\documentclass[amsmath,amssymb,aps,10pt,prd,letterpaper,nofootinbib,balancelastpage,notitlepage,superscriptaddress,floatfix,preprintnumbers]{revtex4-2}
\usepackage{graphicx}	
\usepackage{ragged2e}	
\usepackage{bm}		    
\usepackage{tikz}		
\usetikzlibrary{math}
\usepackage[sort&compress]{natbib}	 	
	\setcitestyle{square,numbers,comma}	
\usepackage[colorlinks=true,urlcolor=blue,linkcolor=red,citecolor=red]{hyperref}	
\newcommand{\figref}[2][]{Fig{#1}.~\ref{fig:#2}}		
\newcommand{\secref}[2][]{Sec{#1}.~\ref{sec:#2}}		
\newcommand{\appref}[2][x]{Appendi{#1}~\ref{app:#2}}	
\renewcommand{\eqref}[2][]{Eq{#1}.~(\ref{eq:#2})}		
\newcommand{\eqrefRange}[2]{Eqs.~(\ref{eq:#1})--(\ref{eq:#2})}		
\newcommand{\citeR}[2][]{Ref{#1}.~\cite{#2}}			
\newcommand{\A}{\hat a}
\newcommand{\B}{\hat b}
\newcommand{\bra}[1]{\langle#1|}

\newcommand{\delrho}{\widehat{\delta\rho}}
\newcommand{\E}{\mathbb E}
\newcommand{\Ex}[2]{\left\langle#1\right\rangle_{#2}}
\newcommand{\ex}[2]{\langle#1\rangle_{#2}}
\let\k\relax
\newcommand{\k}{\mathbf k}
\newcommand{\ket}[1]{|#1\rangle}
\newcommand{\N}{\hat N}
\newcommand{\nl}{\nonumber \\ & \quad }
\newcommand{\p}{\mathbf p}
\newcommand{\q}{\mathbf q}
\newcommand{\x}{\mathbf x}
\newcommand{\X}{\hat X}
\newcommand{\Y}{\hat Y}
\let\Im\relax
\DeclareMathOperator{\Im}{\text{Im}}
\let\Re\relax
\DeclareMathOperator{\Re}{\text{Re}}
\DeclareMathOperator{\Tr}{\text{Tr}}
\DeclareMathOperator{\Var}{\text{Var}}
\DeclareMathSymbol{:}{\mathord}{operators}{"3A}

\begin{document}

\title{Stamps of state on structure:\\
Probing the state of ultralight dark matter via its density fluctuations}
\date{\today}
\author{Saarik Kalia}
\email{kalias@umn.edu}
\affiliation{School of Physics \& Astronomy, University of Minnesota, Minneapolis, MN 55455, USA}

\preprint{UMN-TH-4421/25}

\begin{abstract}
Dark matter (DM) candidates with very small masses, and correspondingly large number densities, have gained significant interest in recent years.  These DM candidates are typically said to behave ``classically".  More specifically, they are often assumed to reside in an ensemble of coherent states.  One notable exception to this scenario is when isocurvature fluctuations of the DM are produced during inflation (or more generally by any Bogoliubov transformation).  In such contexts, the ultralight DM instead resides in a squeezed state.  In this work, we demonstrate that these two scenarios can be distinguished via the statistics of the DM density fluctuations, such as the matter power spectrum and bispectrum.  This provides a probe of the DM state which persists in the limit of large particle number and does not rely on any non-gravitational interactions of the DM.  Importantly, the statistics of these two states differ when the modes of the squeezed state are all in-phase, as is the case at the end of inflation.  Later cosmological dynamics may affect this configuration.  Our work motivates future numerical studies of how cosmological dynamics may impact the initial squeezed state and the statistics of its density fluctuations.
\end{abstract}

\maketitle

\section{Introduction}
\label{sec:introduction}

Copious evidence indicates the existence of non-interacting dark matter (DM) within our Universe~\cite{zwicky1937masses,rubin1970rotation,Clowe_2004,Planck2018_cosmological,descollaboration2025,desicollaboration2025}.  To date, our only observations of DM have been through its gravitational effects on visible matter.  Its fundamental properties, such as its mass and spin, remain unknown, and it is not known whether DM can interact through any means other than gravity.  One class of popular candidates for DM which have gained significant interest in recent years are ultralight candidates with masses $m\lesssim1\,\mathrm{eV}$~\cite{Arias:2012az,Centers:2019dyn,antypas2022,kimball2022search,Cheong_2025}.%
\footnote{We work in natural units $\hbar=c=1$.}
These candidates have high number densities and so must be bosonic in nature.  Popular ultralight bosonic DM candidates include the QCD axion~\cite{Preskill:1982cy,Abbott:1982af,Dine:1982ah}, axionlike particles~\cite{Svrcek:2006yi,Arvanitaki:2009fg,Gra15,co2020predictions}, and dark photons~\cite{Holdom:1986ag,cvetivc1996implications,Nelson:2011sf,Graham:2015rva}.

Owing to their high number densities, ultralight DM candidates are often described as behaving ``classically" in laboratory settings~\cite{kimball2022search,Cheong_2025,kim2023} or when virialized inside the galaxy~\cite{Davidson_2015,lin2018self,Eberhardt:2023axk,kim2024}.  By this, one typically means that its statistics and dynamics can be accurately characterized by a classical field.  As with all fundamental fields, a precise treatment of ultralight DM requires a quantum field description.  In the case of a free real scalar field, the quantum field operator takes the form
\begin{equation}
    \hat\phi(\x,t)=\int\frac{d^3\p}{(2\pi)^3}\frac1{\sqrt{2E_\p}}\left(\A_\p e^{-iE_\p t+i\p\cdot\x}+\A_\p^\dagger e^{iE_\p t-i\p\cdot\x}\right).
    \label{eq:scalar}
\end{equation}
Here $E_\p=\sqrt{|\p|^2+m^2}$, and $\A_\p^\dagger,\A_\p$ represent creation/annihilation operators for particles of momentum $\p$.  In this way, the quantum field consists of a series of quantum harmonic oscillators, associated with each $\p$.

More precisely, when one refers to ultralight DM as ``classical", it is often meant that the state of the DM is described by an ensemble of coherent states with respect to these harmonic oscillators~\cite{Glauber1963,Cheong_2025}.  A pure coherent state is labeled by a set of complex amplitudes $\{\gamma_\k\}$, for each momentum mode $\k$ of the field.  In the limit of large particle number, a quantum field in a coherent state $\ket{\{\gamma_\k\}}$ can be primarily characterized by its expectation value
\begin{equation}
    \phi(\x,t)=\ex{\hat\phi(\x,t)}\gamma\equiv\bra{\{\gamma_\k\}}\hat\phi(\x,t)\ket{\{\gamma_\k\}},
\end{equation}
which behaves as a classical field (e.g., it obeys the Klein-Gordon equation).  The DM resides in a mixed state formed by a classical ensemble of these coherent states.  We will represent such a state by its density matrix $\Psi$.  Certain production mechanisms, such as misalignment production~\cite{Preskill:1982cy,Dine:1982ah,Svrcek:2006yi,Nelson:2011sf,Davidson_2015,Graham_2018,Arvanitaki_2020}, may already produce the DM in a coherent state ensemble.  Even if the DM does not originate in this state, it is often argued that galactic virialization will generically place it in a coherent state ensemble~\cite{lin2018self,Eberhardt:2023axk}, and so this description is appropriate for studying dark matter on astrophysical or terrestrial scales.

One notable context in which DM fluctuations are not produced in a coherent state is inflationary production~\cite{Ford1987,Mukhanov:1990me,Lyth_1998,Graham:2015rva}.  When the Universe undergoes a period of accelerated expansion, the vacuum state after the expansion differs from the vacuum state in the asymptotic past.  This is captured by a Bogoliubov transformation of the creation and annihilation operators
\begin{align}\label{eq:bogoliubov1}
    \A_\p&=\alpha_\p\B_\p+\beta_\p\B_{-\p}^\dagger,\\
    \A_\p^\dagger&=\alpha_\p^*\B_\p^\dagger+\beta_\p^*\B_{-\p}.
    \label{eq:bogoliubov2}
\end{align}
Here $\B_\p,\B^\dagger_\p$ represent the creation/annihilation operators in the asymptotic past, while $\A_\p,\A_\p^\dagger$ represent the operators after the expansion.%
\footnote{Note that the roles of $\A_\p$ and $\B_\p$ in our work are opposite to the convention often used in the literature for a Bogoliubov transformation.  We also note that, in the quantum optics literature, the Bogoliubov coefficients are sometimes expressed in terms of a squeezing parameter $r$ and phases $\theta_1,\theta_2$ as $\alpha=e^{i\theta_1}\cosh r,\beta=e^{i\theta_2}\sinh r$.}
The DM begins in the vacuum state $\ket\Omega$ of $\B_\p,\B_\p^\dagger$.  After the expansion, $\ket\Omega$ is not the vacuum of the new operators $\A_\p,\A_\p^\dagger$, and so this implies that particle production has occurred.  Note that the state $\ket\Omega$ is also not a coherent state of $\A_\p,\A_\p^\dagger$, but rather it is a squeezed state~\cite{Albrecht_1994,Polarski_1996,Ku_2021}.  More specifically, in the context of inflation, the modes of the scalar field become frozen when they exit the horizon, so that $\partial_t\hat\phi$ is very small for each mode, i.e., the state is squeezed along the orthogonal direction.  Inflationary production (or any mechanism which produces DM fluctuations via a Bogoliubov transformation) may therefore result in fluctuations on cosmological scales which are not described by a coherent state ensemble.

If the DM is described by an ultralight bosonic field, such fluctuations will necessarily be produced during inflation.  Most notably, in the case of vector DM, inflation can produce the entirety of DM, without violating isocurvature constraints~\cite{Graham:2015rva}.  In the case of a scalar field, inflationary production leads to isocurvature fluctuations on large scales~\cite{fox2004,Hamann_2009}, which are constrained by observations of the cosmic microwave background (CMB) and large-scale structure (LSS)~\cite{Planck2018_inflation,buckley2025}.  Because LSS alone cannot probe isocurvature, these limits can only directly constrain scales $\gtrsim10\,\mathrm{Mpc}$, while lower scales require an assumption on the spectral shape of isocurvature fluctuations~\cite{buckley2025}.  In this work, we will remain agnostic to this spectral shape.

The primary purpose of this work is to demonstrate that if isocurvature fluctuations of the DM are produced during inflation (or more generally, through any Bogoliubov transformation), they may exist in a different state than the usual coherent state ensemble which is assumed for ultralight DM.  Moreover, we show that these two different scenarios can be distinguished solely by comparing the statistics of the density perturbations of DM.  This distinction relies on no non-gravitational interactions of the DM, and can probe smaller scales than direct isocurvature constraints.  For these reasons, the matter bispectrum and higher-point statistics may be powerful probes of the DM's current state and of its origin~\cite{Geller:2024upd}.

This work is organized as follows.  In \secref{qho}, we review the definitions of and distinction between a coherent state ensemble and squeezed state, in the context of a quantum harmonic oscillator.  We demonstrate that these two states exhibit different statistics for the number operator, so that they can be distinguished solely by measuring particle number.  In \secref{field}, we define these two states in the context of a quantum field.  In this work, we focus on a real scalar field, as defined in \eqref{scalar}, but our results can be generalized to vector fields as well.  In \secref{perturbations}, we demonstrate that the statistics of the DM density perturbations differ between a coherent state ensemble and a squeezed state (in analogy to the number operator in the quantum harmonic oscillator).  Importantly, this distinction persists in the limit of large particle number.  Finally, in \secref{discussion}, we conclude.

Throughout this work, we will assume that the current state of DM is a coherent state ensemble or squeezed state, without considering how its state might be altered by cosmological dynamics.  In \appref{dynamics}, we address this point, and show that if the DM begins in a squeezed state, the dynamics of its state may be nontrivial.  In \appref{4point}, we show that the four-point statistics of the density perturbations may be useful in the case where the squeezed state is disrupted.  We leave a more detailed treatment of these dynamics and how it affects the statistics of the density perturbations to future work.

\section{Quantum harmonic oscillator}
\label{sec:qho}

We begin by considering a quantum harmonic oscillator with frequency $\omega$, described by the Hamiltonian
\begin{equation}
    \hat H=\omega\left(\A^\dagger\A+\frac12\right)=\omega\left(\X^2+\Y^2\right).
    \label{eq:hamiltonian}
\end{equation}
This Hamiltonian can either be expressed in terms of the annihilation/creation operators $\A,\A^\dagger$, which satisfy the canonical commutation relation $[\A,\A^\dagger]=1$, or in terms of the related quadrature operators
\begin{equation}
    \X=\frac{\A+\A^\dagger}2,\qquad\Y=-\frac{i(\A-\A^\dagger)}2
\end{equation}
(which are proportional to the usual position and momentum operators).

\begin{figure*}[t]
\includegraphics[width=0.49\textwidth]{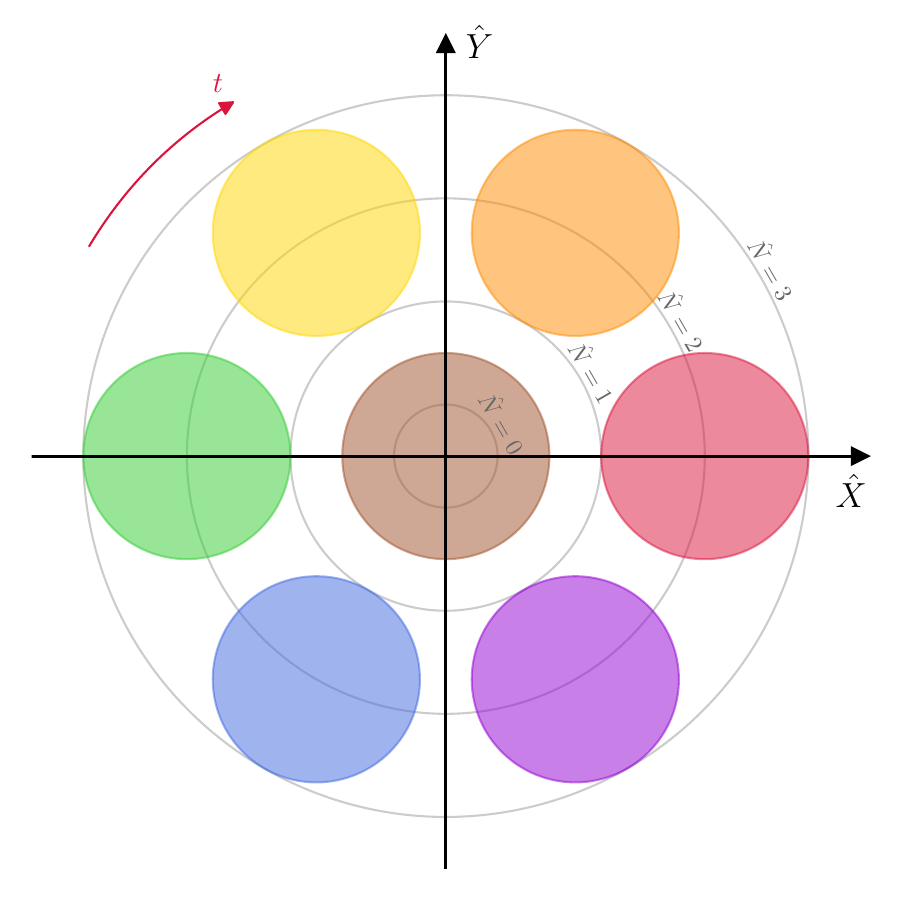}
\includegraphics[width=0.49\textwidth]{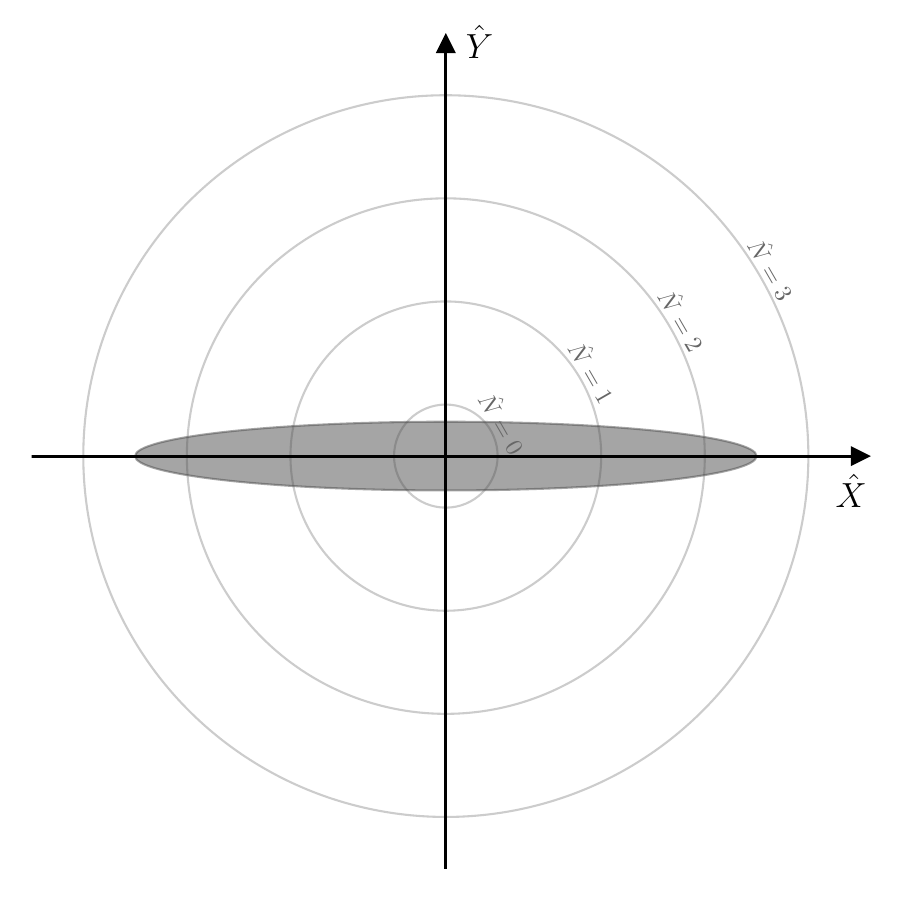}
\caption{\label{fig:states}%
    Schematic diagram of an ensemble of coherent states (left) and a squeezed state (right) in the phase space parametrized by $\X$ and $\Y$.  Each circle/ellipse represents the $2\sigma$-contour of the probability density of a state.  On the left, each colored circle represents a single coherent state, with the brown state being the vacuum.  The center of each circle lies at $(\X,\Y)=(\Re[\gamma],\Im[\gamma])$.  A Gaussian coherent state ensemble is a classical ensemble of such states, with $\gamma$ distributed as in \eqref{probability}.  On the right, the grey ellipse represents a state which is squeezed along the $\X$-direction.  Time evolution in these phase space diagrams corresponds to clockwise rotation.  Note that the coherent state ensemble is time-invariant, while the squeezed state is not.  In light grey, we show several contours of fixed $\N=\X^2+\Y^2-\frac12$.  It is apparent that these two states exhibit different statistics for $\N$, that is if both are scaled to have the same expectation of $\N$, the squeezed state will have a larger variance than the coherent state ensemble.  In the case of a quantum field, the roles of $\hat X$ and $\hat Y$ are played by (the momentum modes of) $\hat\phi$ and $\partial_t\hat\phi$, respectively.  Importantly, inflation produces a squeezed state where all modes are squeezed along the $\hat\phi$-direction [see discussion following \eqref{2point_squeezed}].}
\end{figure*}

It is not difficult to show that the Hamiltonian in \eqref{hamiltonian} implies that these operators evolve (in the Heisenberg picture) as
\begin{align}
    \A(t)&=\A(0)e^{-i\omega t},\\
    \A^\dagger(t)&=\A^\dagger(0)e^{i\omega t},\\
    \X(t)&=\frac{\A(0)e^{-i\omega t}+\A^\dagger(0)e^{i\omega t}}2\label{eq:Xt}\\
    &=\X(0)\cos\omega t+\Y(0)\sin\omega t,\\
    \Y(t)&=\Y(0)\cos\omega t-\X(0)\sin\omega t.
\end{align}
In \figref{states}, we show various states in the phase space parameterized by $\X$ and $\Y$.  (This illustration can be more precisely formalized in terms of the Wigner function~\cite{Wigner1932}.)  Time evolution corresponds to clockwise rotation in this phase space.

We will consider two different states of this oscillator.  The first is a classical Gaussian ensemble of coherent states.  This is the state which most closely corresponds to Gaussian noise in a classical harmonic oscillator.  As it is a classical ensemble of states, it will be described by a density matrix.  The second state will be constructed by beginning with the vacuum state of the oscillator and then performing a Bogoliubov transformation.  As we will see, this will lead to a squeezed state, i.e. its variance will be smaller along one quadrature and larger along the other.  We will consider the statistics of the number operator $\N=\A^\dagger\A$ in both of these states and find that the squeezed state exhibits a larger variance than the ensemble of coherent states (when their expectations are equal).

\subsection{Ensemble of coherent states}
\label{sec:qho_coherent}

A coherent state is defined as an eigenvector of the annihilation operator, i.e., $\A\ket\gamma=\gamma\ket\gamma$.  As can be seen in \figref{states}, this state is localized in phase space around its expectation values
\begin{align}
    \ex\X\gamma&\equiv\bra\gamma\X\ket\gamma=\Re[\gamma],\\
    \ex\Y\gamma&=\Im[\gamma].
\end{align}
As such, it mimics the dynamics of a classical harmonic oscillator (which would correspond to a definite point in phase space).  The number operator in this state has expectation and variance
\begin{align}
    \ex\N\gamma&=|\gamma|^2,\\
    \Var_\gamma(\N)&\equiv\ex{\N^2}\gamma-\ex\N\gamma^2=|\gamma|^2.
\end{align}
A single coherent state satisfies $\Var_\gamma(\N)=\ex\N\gamma$.  This describes the quantum uncertainty of a coherent state.  Its relative uncertainty vanishes in the large number limit, in the sense that
\begin{equation}
    \frac{\sqrt{\Var_\gamma(N)}}{\ex\N\gamma}\rightarrow0
\end{equation}
as $|\gamma|\rightarrow\infty$.

A single coherent state corresponds to a fixed classical realization of a harmonic oscillator.  We will be interested in the statistics of this system, and so we should instead consider a classical ensemble of these coherent states.  Such an ensemble is defined by a density matrix%
\footnote{Here $\int d\gamma=\int d(\Re[\gamma])\cdot d(\Im[\gamma])$ represents a double integral over the real and imaginary parts of $\gamma$.}
\begin{equation}
    \Psi=\int d\gamma\,p(\gamma)\ket\gamma\bra\gamma,
\end{equation}
where $p(\gamma)$ is a probability distribution describing the classical ensemble.  In this work, we will consider a Gaussian distribution
\begin{equation}
    p(\gamma)=\frac1{\pi\sigma^2}e^{-\frac{|\gamma|^2}{\sigma^2}},
    \label{eq:probability}
\end{equation}
with some width $\sigma$, as has been considered in \citeR[s]{kim2023,kim2024,Cheong_2025}.  This probability distribution has unit norm, so that $\Tr[\Psi]=1$.  The statistics of the number operator in this ensemble are
\begin{align}
    \ex\N\Psi&\equiv\Tr[\Psi\N]=\E[|\gamma|^2]=\sigma^2,\\
    \ex{\N^2}\Psi&=\E[|\gamma|^4+|\gamma|^2]=2\sigma^4+\sigma^2,
    \label{eq:N2_coherent}
\end{align}
where $\E[f(\gamma)]\equiv\int d\gamma\,p(\gamma)f(\gamma)$ implies an expectation over the classical probability distribution.  Therefore, a Gaussian ensemble of coherent states satisfies
\begin{equation}
    \Var_\Psi(\N)=\ex\N\Psi^2+\ex\N\Psi.
\end{equation}
The latter contribution is the same quantum uncertainty we found for a single coherent state.  The former contribution represents the classical uncertainty from the Gaussian ensemble.  In the large number limit, the classical uncertainty dominates, i.e.
\begin{equation}
    \frac{\sqrt{\Var_\Psi(N)}}{\ex\N\Psi}\rightarrow1
    \label{eq:largeN_Psi}
\end{equation}
as $\sigma\rightarrow\infty$.

\subsection{Squeezed state}
\label{sec:qho_squeezed}

Now let us consider the state $\ket\Omega$ which is annihilated by a different annihilation operator $\B$.  Let the operators $\B,\B^\dagger$ be related to $\A,\A^\dagger$ by a Bogoliubov transformation
\begin{align}
    \A&=\alpha\B+\beta\B^\dagger,\\
    \A^\dagger&=\alpha^*\B^\dagger+\beta^*\B.
\end{align}
The coefficients must satisfy $|\alpha|^2-|\beta|^2=1$, in order for both $\A$ and $\B$ to satisfy their canonical commutation relations.  The state $\ket\Omega$ is the vacuum state of the quantum harmonic oscillator defined by $\B$.  For the harmonic oscillator defined by $\A$, it is instead a squeezed state.  This can be seen by considering the quadrature operators $\X$ and $\Y$.  The expectations of both of these operators are zero in the state $\ket\Omega$.  However, if we consider instead the variance of a linear combination of the quadrature operators, we find
\begin{align}
    \Var_\Omega(\cos\theta\cdot\X+\sin\theta\cdot\Y)&=\Var_\Omega\left(\frac{e^{i\theta}\A^\dagger+e^{-i\theta}\A}2\right)=\Ex{\left(\frac{\left(e^{i\theta}\alpha^*+e^{-i\theta}\beta\right)\B^\dagger+\left(e^{i\theta}\beta^*+e^{-i\theta}\alpha\right)\B}2\right)^2}\Omega\\
    &=\frac{\left|e^{i\theta}\beta^*+e^{-i\theta}\alpha\right|^2}4=\frac{|\alpha|^2+|\beta|^2+2\Re[e^{-2i\theta}\alpha\beta]}4.
\end{align}
We see that the variance is minimized for $\theta=\frac{\pi+\arg[\alpha\beta]}2$, in which case it becomes $\frac{(|\alpha|-|\beta|)^2}4$.  On the other hand, the maximum variance $\frac{(|\alpha|+|\beta|)^2}4$ is achieved for $\theta=\frac{\arg[\alpha\beta]}2$.  In this sense, the state has been ``squeezed" along the axes corresponding to these choices of $\theta$.  Note that in accordance with the uncertainty principle, the product of the variances along any two orthogonal axes is at least $\frac1{16}$.  In \figref{states}, we show a state $\ket\Omega$ which is squeezed along the $\theta=0$ direction (i.e. along the $\X$-axis).

Now let us compute the statistics of the number operator (corresponding to $\A$) in this squeezed state
\begin{align}
    \ex\N\Omega&=\Ex{\left(\alpha^*\B^\dagger+\beta^*\B\right)\left(\alpha\B+\beta\B^\dagger\right)}\Omega=|\beta|^2,\\
    \ex{\N^2}\Omega&=|\alpha|^2|\beta|^2\ex{\B\B\B^\dagger\B^\dagger}\Omega+|\beta|^4\ex{\B\B^\dagger\B\B^\dagger}\Omega=2|\alpha|^2|\beta|^2+|\beta|^4=3|\beta|^4+2|\beta|^2.
    \label{eq:N2_squeezed}
\end{align}
A squeezed state therefore satisfies
\begin{equation}
    \Var_\Omega(\N)=2\ex\N\Omega^2+2\ex\N\Omega.
\end{equation}
We see that the number operator has twice the variance in a squeezed state as it does in an ensemble of coherent states (if its expectation is held fixed)!  Importantly, this is an effect which persists in the large number limit, i.e.
\begin{equation}
    \frac{\sqrt{\Var_\Omega(N)}}{\ex\N\Omega}\rightarrow\sqrt2
    \label{eq:largeN_Omega}
\end{equation}
as $|\beta|\rightarrow\infty$.  Examining the statistics of $\hat N$ therefore offers a way to distinguish a coherent state ensemble from a squeezed state, even in the limit of many particles.  We will apply this same logic to a quantum field, and identify a corresponding observable that can distinguish ultralight DM in a coherent state ensemble, as is typically presumed, from DM in a squeezed state, which could be the case if it were produced via a Bogoliubov transformation.

Before proceeding, we comment on the implications of these results on the classicality of these states.  A coherent state is typically associated with ``classical" behavior of a quantum harmonic oscillator, and so one may be tempted to infer that a squeezed state is therefore not classical.  Comparing \eqref{largeN_Psi} to \eqref{largeN_Omega} would then offer a way to discern the ``classical" or ``quantum" nature of the state, even in the large number limit.  We stress that both coherent states and squeezed states may be interpretted as classical, in the large number limit, and so we do not claim that the comparison made here relates to the quantum nature of the state.  A coherent state ensemble acts classically because its dynamics mimic an ensemble of classical trajectories with initial state drawn from \eqref{probability} [in the limit $\sigma\rightarrow\infty$, where the finite size of each state in phase space can be ignored].  Likewise, in the limit $|\beta|\rightarrow\infty$, the dynamics of a state which is squeezed along the $\X$-direction can be fully described by an ensemble of classical trajectories with initial $X$ drawn from a Gaussian distribution of width $|\beta|$ and initial $Y=0$.  In this sense, a squeezed state corresponds to a classical state with \emph{one} degree of freedom, while a coherent state ensemble corresponds to a classical state with \emph{two} degrees of freedom.  Nevertheless, they can be distinguished solely through the statistics of $\N$.

\section{Quantum field}
\label{sec:field}

In this section, we define the notions of coherent and squeezed states in the context of a quantum scalar field.%
\footnote{Many of the results in this section can be generalized to massive vector fields.  Each polarization of the vector field can be separately quantized, as we do for scalar fields.  In general, care must be taken when treating the longitudinal vs.~transverse modes of the field.  Since we consider DM in this work, we ultimately take the non-relativistic limit [see \eqref{delrho}], in which case the distinction between the longitudinal and transverse modes becomes negligible.  The final results will then only differ from the scalar case by $\mathcal O(1)$ factors.}
A free scalar field can be decomposed into momentum modes, each of which behaves as an independent quantum harmonic oscillator, as in \eqref{scalar}.  The creation and annihilation operators in this expression satisfy the commutation relation
\begin{equation}
    [\A_\p,\A_\q^\dagger]=(2\pi)^3\delta^{(3)}(\p-\q).
    \label{eq:commutation_field}
\end{equation}
We will define coherent and squeezed states similarly to how we did in \secref{qho}, only now we will need to specify the state of each oscillator.  We can see from comparing \eqref{scalar} to \eqref{Xt} that (the momentum modes of) $\hat\phi$ play an analogous role to $\X$, while its derivative $\partial_t\hat\phi$ is analogous to $\Y$.  We can therefore understand these states in terms of the phase space in \figref{states}, just as we did for the quantum harmonic oscillator.

We will also derive the statistics of the creation and annihilation operators in a coherent state ensemble vs. a squeezed state.  We will see that while in the coherent state ensemble, only contractions of $\A_\p$ with $\A_\q^\dagger$ are nonzero, in the squeezed state $\A_\p$ and $\A_\q$ will have nonzero contractions (and likewise for $\A_\p^\dagger$ and $\A_\q^\dagger$).  This key finding will allow us to evaluate more complicated correlators in \secref{perturbations}.

\subsection{Ensemble of coherent states}
\label{sec:field_coherent}

A coherent state of $\hat\phi$ is defined in a similar way to a coherent state in a quantum harmonic oscillator.  In this case, we describe a coherent state by a set of complex numbers $\{\gamma_\k\}$ for each momentum mode $\k$, so that the associated coherent state is defined as
\begin{equation}
    \A_\p\ket{\{\gamma_\k\}}=\gamma_\p\ket{\{\gamma_\k\}}.
\end{equation}
The expectation of the scalar field in this state is then
\begin{equation}
    \ex{\hat\phi(\x,0)}{\{\gamma_\k\}}\equiv\bra{\{\gamma_\k\}}\hat\phi(\x,0)\ket{\{\gamma_\k\}}=\int\frac{d^3\p}{(2\pi)^3}\left(\frac{\gamma_\p+\gamma_{-\p}^*}{\sqrt{2E_\p}}\right)e^{i\p\cdot\x}\equiv\int\frac{d^3\p}{(2\pi)^3}\phi_\p e^{i\p\cdot\x}.
\end{equation}
Similarly, the conjugate field has expectation
\begin{equation}
    \ex{\partial_t\hat\phi(\x,0)}{\{\gamma_\k\}}=\int\frac{d^3\p}{(2\pi)^3}\left(-i\sqrt{\frac{E_\p}2}\left(\gamma_\p-\gamma_{-\p}^*\right)\right)e^{i\p\cdot\x}\equiv\int\frac{d^3\p}{(2\pi)^3}\pi_\p e^{i\p\cdot\x}.
\end{equation}
We then see that a coherent state may be equivalently described by the sets of complex numbers $\{\phi_\k,\pi_\k\}$ (which must satisfy $\phi_{-\k}=\phi_\k^*$ and $\pi_{-\k}=\pi_\k^*$) instead of the numbers $\{\gamma_\k\}$.

Again we wish to consider a Gaussian ensemble of coherent states.  As the ensemble is Gaussian, it is completely defined by its two-point statistics.  The most useful quantity to define these statistics is the power spectrum $P(\p)$, which determines the two-point statistics of $\phi_\p$ and $\pi_\p$,
\begin{align}\label{eq:expec_phi1}
    \E[\phi_\p\phi_\q^*]&=(2\pi)^3P(\p)\delta^{(3)}(\p-\q),\\
    \E[\pi_\p\pi_\q^*]&=(2\pi)^3E_\p^2P(\p)\delta^{(3)}(\p-\q),\\
    \E[\phi_\p\pi_\q^*]&=0.
\end{align}
Note that \eqref{expec_phi1} and $\phi_{-\p}=\phi_\p^*$ implies $P(-\p)=P(\p)$.  The two-points statistics of $\gamma_\p$ are then
\begin{align}\label{eq:expec_gamma1}
    \E[\gamma_\p\gamma_\q^*]&=(2\pi)^3E_\p P(\p)\delta^{(3)}(\p-\q),\\
    \E[\gamma_\p\gamma_\q]&=\E[\gamma_\p^*\gamma_\q^*]=0.
    \label{eq:expec_gamma2}
\end{align}
If $\Psi$ is the density matrix for the Gaussian ensemble, then \eqref[s]{expec_gamma1} and (\ref{eq:expec_gamma2}) imply the operator traces
\begin{align}\label{eq:trace_a1}
    \ex{\A_\p^\dagger\A_\q}\Psi&\equiv\Tr[\Psi\A_\p^\dagger\A_\q]=(2\pi)^3E_\p P(\p)\delta^{(3)}(\p-\q),\\
    \ex{\A_\p\A_\q^\dagger}\Psi&=(2\pi)^3(E_\p P(\p)+1)\delta^{(3)}(\p-\q)\label{eq:trace_a2},\\
    \ex{\A_\p\A_\q}\Psi&=\ex{\A_\p^\dagger\A_q^\dagger}\Psi=0.
    \label{eq:trace_a3}
\end{align}

Higher-point statistics in a Gaussian ensemble can be computed via Wick's theorem, that is, any expectation of a product of $\gamma_\p$'s and $\gamma_\p^*$'s can be computed by considering all possible contractions of the product and applying the two-point statistics in  \eqref[s]{expec_gamma1} and (\ref{eq:expec_gamma2}).  Likewise, any trace of a product of creation and annihilation operators can be computed by applying \eqrefRange{trace_a1}{trace_a3} to all possible contractions (paying careful attention to the relative ordering of any contracted pair).  Note that in this context, we need only sum over \emph{bipartite} contractions, that is, contractions which pair each creation operator with an annihilation operator.  \eqref{trace_a3} implies that any contraction which pairs two creation or two annihilation operators will vanish.  As an example, we compute the operator trace
\begin{align}\label{eq:trace_4as1}
    \ex{\A_{\p_1}^\dagger\A_{\p_2}\A_{\p_3}^\dagger\A_{\p_4}}\Psi&=\ex{\A_{\p_1}^\dagger\A_{\p_2}}\Psi\ex{\A_{\p_3}^\dagger\A_{\p_4}}\Psi+\ex{\A_{\p_1}^\dagger\A_{\p_4}}\Psi\ex{\A_{\p_2}\A_{\p_3}^\dagger}\Psi\\
    &=(2\pi)^6E_{\p_1} P(\p_1)E_{\p_3} P(\p_3)\delta^{(3)}(\p_1-\p_2)\delta^{(3)}(\p_3-\p_4)\nl
    +(2\pi)^6E_{\p_1} P(\p_1)(E_{\p_3} P(\p_3)+1)\delta^{(3)}(\p_1-\p_4)\delta^{(3)}(\p_2-\p_3).
    \label{eq:trace_4as2}
\end{align}

\subsection{Squeezed state}
\label{sec:field_squeezed}

Now we move on to consider a squeezed state for the field $\hat\phi$.  Again we define the state $\ket\Omega$ as the vacuum with respect to a different set of annihilation operators $\B_\p$.  These operators are related to the $\A_\p$ from which we construct $\hat\phi$ via a Bogoliubov transformation, as in \eqref[s]{bogoliubov1} and (\ref{eq:bogoliubov2}), with Bogoliubov coefficients $\alpha_\p,\beta_\p$ for each momentum mode.  In order for both $\A_\p$ and $\B_\p$ to satisfy \eqref{commutation_field}, again these must satisfy $|\alpha_\p|^2-|\beta_\p|^2=1$, but now in addition we also require $\alpha_\p\beta_{-\p}=\alpha_{-\p}\beta_\p$.  Note that in order to satisfy momentum conservation, $\A_\p$ is a linear combination of $\B_\p$ and $\B_{-\p}^\dagger$ (and likewise for $\A_\p^\dagger$).  With these definitions, we find the two-point expectation values
\begin{align}\label{eq:expec_sq1}
    \ex{\A_\p^\dagger\A_\q}\Omega&\equiv\bra\Omega\A_\p^\dagger\A_\q\ket\Omega=(2\pi)^3|\beta_\p|^2\delta^{(3)}(\p-\q),\\
    \ex{\A_\p\A_\q^\dagger}\Omega&=(2\pi)^3|\alpha_\p|^2\delta^{(3)}(\p-\q)\label{eq:expec_sq2},\\
    \ex{\A_\p\A_\q}\Omega&=(2\pi)^3\alpha_\p\beta_{-\p}\delta^{(3)}(\p+\q)\label{eq:expec_sq3},\\
    \ex{\A_\p^\dagger\A_\q^\dagger}\Omega&=(2\pi)^3\alpha_{-\p}^*\beta_\p^*\delta^{(3)}(\p+\q).
    \label{eq:expec_sq4}
\end{align}

We see that \eqref[s]{expec_sq1} and (\ref{eq:expec_sq2}) take a very similar form to \eqref[s]{trace_a1} and (\ref{eq:trace_a2}).  In general, let us define the power spectrum $P(\p)$ for any state by \eqref{trace_a1}.  The squeezed state then satisfies $|\beta_\p|^2=E_\p P(\p)$.  \eqref[s]{expec_sq3} and (\ref{eq:expec_sq4}), however, demonstrate one notable difference between a squeezed state and a coherent state ensemble: the product of two creation or two annihilation operators has a nonzero expectation value in a squeezed state!  This is a crucial observation for computing higher-point expectation values.  Whereas for the coherent state ensemble, we applied Wick's theorem over all bipartite contractions, for the squeezed state, we must sum over \emph{all} contractions (including ones between two creation operators or between two annihilation operators).  As an example, we can compute the expectation value of the operator in \eqref{trace_4as1} in a squeezed state
\begin{align}\label{expec_4as1}
    \ex{\A_{\p_1}^\dagger\A_{\p_2}\A_{\p_3}^\dagger\A_{\p_4}}\Omega&=\ex{\A_{\p_1}^\dagger\A_{\p_2}}\Omega\ex{\A_{\p_3}^\dagger\A_{\p_4}}\Omega+\ex{\A_{\p_1}^\dagger\A_{\p_4}}\Omega\ex{\A_{\p_2}\A_{\p_3}^\dagger}\Omega+\ex{\A_{\p_1}^\dagger\A_{\p_3}^\dagger}\Omega\ex{\A_{\p_2}\A_{\p_4}}\Omega\\
    &=(2\pi)^6E_{\p_1} P(\p_1)E_{\p_3} P(\p_3)\delta^{(3)}(\p_1-\p_2)\delta^{(3)}(\p_3-\p_4)\nl
    +(2\pi)^6E_{\p_1} P(\p_1)(E_{\p_3} P(\p_3)+1)\delta^{(3)}(\p_1-\p_4)\delta^{(3)}(\p_2-\p_3)\nl
    +(2\pi)^6\alpha_{-\p_1}^*\alpha_{\p_2}\beta_{\p_1}^*\beta_{-\p_2}\delta^{(3)}(\p_1+\p_3)\delta^{(3)}(\p_2+\p_4).
    \label{eq:expec_4as2}
\end{align}
While the first two terms in \eqref{expec_4as2} match \eqref{trace_4as2}, the squeezed state expectation value receives an additional contribution from the contraction of like operators.  This is analogous to the factor of three in \eqref{N2_squeezed} compared to the corresponding factor of two in \eqref{N2_coherent}.  Note that this final term cannot be naively interpreted in terms of the power spectrum, without some additional assumption on the phases of $\alpha_\p$ and $\beta_\p$.  We will return to this point later.

\section{Density perturbations}
\label{sec:perturbations}

In \secref{qho}, we saw that we could distinguish an ensemble of coherent states from a squeezed state by examining the statistics of the number operator $\N$.  In this section, we will attempt to find an analogous quantity in the context of a quantum scalar field.  The primary observables that we can infer from non-interacting DM are its density perturbations.  Here we will show that these are sufficient to distinguish a coherent state ensemble from a squeezed state.  Namely, we will compute various statistics of the density perturbation operator, and explicitly show that they differ between these two states.  In this sense, we can learn about the state of ultralight DM without relying on any non-gravitational interactions.

The energy density operator is defined in terms of the scalar field $\hat\phi$ and can be decomposed into perturbation modes as
\begin{equation}
    \hat\rho(\x,t)=\frac12\left(\partial_t\hat\phi(\x,t)\right)^2+\frac12\left|\nabla\hat\phi(\x,t)\right|^2+\frac12m^2\hat\phi(\x,t)^2\equiv\int\frac{d^3\p}{(2\pi)^3}\delrho(\p,t)e^{i\p\cdot\x}.
\end{equation}
From \eqref{scalar}, we can write the density perturbation operator in terms of the creation and annihilation operators
\begin{align}
    \delrho(\p,t)=\int&\frac{d^3\k}{(2\pi)^3}\frac1{4\sqrt{E_\k E_{\k-\p}}}\Bigg[\left(E_\k E_{\k-\p}+\k\cdot(\k-\p)+m^2\right)\left(\A_\k\A_{\k-\p}^\dagger+\A_{\k-\p}^\dagger\A_\k\right)e^{-i(E_\k-E_{\k-\p})t}\nl
    +\left(-E_\k E_{\k-\p}+\k\cdot(\k-\p)+m^2\right)\left(\A_\k\A_{-\k+\p}e^{-i(E_\k+E_{\k-\p})t}+\A_{\k-\p}^\dagger\A_{-\k}^\dagger e^{i(E_\k+E_{\k-\p})t}\right)\Bigg].
    \label{eq:delrho_full}
\end{align}

In this work, we will primarily be interested in $\delrho(\p,t)$ for $\p\neq0$.  This is because we only have one realization of the total energy over all space $\hat E=\delrho(0,t)$, and so correlators involving this quantity can not be observationally measured.  Correlators involving $\delrho(\p,t)$ for $\p\neq0$, on the other hand, can be measured by observing momentum modes with the same magnitude but different direction (as long as we assume isotropy).  Throughout, we will assume $\p\neq0$, and so in particular, this implies that $\A_\k\A_{\k-\p}^\dagger=\A_{\k-\p}^\dagger\A_\k$ in the first line of \eqref{delrho_full}.  As we are primarily interested in DM, we will also consider the non-relativistic limit, where this integral is dominated by $|\k|\ll m$ (and we consider $|\p|\ll m$).  In this case, \eqref{delrho_full} simplifies to
\begin{equation}
    \delrho(\p,t)=m\int\frac{d^3\k}{(2\pi)^3}\A_{\k-\p}^\dagger\A_\k.
    \label{eq:delrho}
\end{equation}
Note that in this limit, the time-dependence of $\delrho(\p,t)$ vanishes (see \appref{dynamics} for further discussion of time-dependence), and so we will suppress the explicit time-dependence in our notation henceforth.  Finally, to remove explicit UV divergences, we will consider normally ordered correlators in this section.  We will evaluate the two- and three-point statistics of $\delrho(\p)$ in a coherent vs. a squeezed state, just as we did for the number operator in \secref{qho}.

\subsection{Two-point function}
\label{sec:2point}

We begin by computing the normally ordered equal-time two-point function of the density perturbation operator in both of the states of interest.  Let us begin with the coherent state ensemble.  As discussed in \secref{field}, we can compute the two-point function by applying Wick's theorem over all bipartite contractions of creation and annihilation operators.  For two density perturbation operators $\delrho(\p),\delrho(\q)$, as in \eqref{delrho}, there are naively two such contractions.  However because we are interested in $\p,\q\neq0$, only one of these is nonzero.  This yields the normally ordered two-point function%
\footnote{\label{ftnt:Pk_behavior}%
Depending on the large-$|\k|$ behavior of $P(\k)$, the integral in \eqref{2point} may naively be ultraviolet (UV) dominated, so that relativistic modes should contribute.  We note that in order for the mean energy density [see \eqref{rhobar}] to be dominated by non-relativistic modes, $P(\k)$ must already fall off faster than $|\k|^{-3}$.  This scaling also makes \eqref{2point} UV finite, and so justifies the non-relativistic assumption.}
\begin{equation}
    \ex{:\delrho(\p)\delrho(\q):}\Psi=m^2\int\frac{d^3\k}{(2\pi)^3}\frac{d^3\k'}{(2\pi)^3}\ex{\A^\dagger_{\k-\p}\A_{\k'}}\Psi\ex{\A^\dagger_{\k'-\q}\A_\k}\Psi=\delta^{(3)}(\p+\q)\cdot m^4\int d^3\k\,P(\k)P(\k-\p).
    \label{eq:2point}
\end{equation}

Let us now consider the two-point function in a squeezed state.  As in the coherent state ensemble, we will normal order the two-point density perturbation operator (with respect to the $\A_\p,\A_\p^\dagger$ operators, not the $\B_\p,\B_\p^\dagger$ operators) before taking its expectation in the $\ket\Omega$ state.  As discussed in \secref{field}, the crucial difference between the coherent state ensemble and the squeezed state is that non-bipartite contractions contribute to the latter.  This gives one additional contribution to the squeezed state which did not appear in the coherent state ensemble
\begin{align}
    \ex{:\delrho(\p)\delrho(\q):}\Omega-\ex{:\delrho(\p)\delrho(\q):}\Psi&=m^2\int\frac{d^3\k}{(2\pi)^3}\frac{d^3\k'}{(2\pi)^3}\ex{\A_\k\A_{\k'}}\Omega\ex{\A_{\k-\p}^\dagger\A_{\k'-\q}^\dagger}\Omega\\
    &=\delta^{(3)}(\p+\q)\cdot m^2\int d^3\k\,\alpha_\k\beta_{-\k}\alpha_{\p-\k}^*\beta_{\k-\p}^*.
    \label{eq:2point_squeezed}
\end{align}

At this point, we must make some assumptions about the phases of the Bogoliubov coefficients $\alpha_\p,\beta_\p$, which generically may depend on how the squeezed state is produced.  If the phases of the Bogoliubov coefficients vary as a function of momentum, the integral in \eqref{2point_squeezed} may be highly oscillatory, so that the two-point function for the squeezed state reduces to the case of the coherent state ensemble.  However, if the squeezed state is produced during inflation, the phases of the Bogoliubov coefficients will (at least initially) be independent of $\p$~\cite{Ku_2021}!  This is because, during inflation, momentum modes become frozen once they exit the horizon, i.e. $\partial_t\hat\phi$ is very small for all (sufficiently small) $\p$ at the end of inflation.  In the language of squeezing, this implies that $\arg[\alpha_\p\beta_{-\p}]=0$ for all such $\p$, so that all modes are squeezed along the $\hat\phi$-direction (in analogy to the squeezing in the $\X$-direction shown in \figref{states}).  These modes re-enter the horizon when the Hubble constant equals the mass of the field.%
\footnote{This is only true for modes which re-enter while non-relativistic.  Modes which re-enter while relativistic will do so earlier, and thus will accumulate some additional phase relative to non-relativistic modes.  This phase is subdominant to later dispersion which occurs after all modes have re-entered [see \eqref{dispersion_int}].}
From then on, the phases of all modes evolve with frequency $E_\p\approx m$ (represented by clockwise evolution in \figref{states}).  Importantly because all modes begin this evolution at roughly the same time, they will remain in-phase until dispersion causes their relative phases to drift.  For the remainder of this section, we will compute the two- and three-point functions of $\delrho(\p)$ under the assumption that the Bogoliubov coefficients remain in-phase.  In \appref{dynamics}, we estimate the impact of dynamical effects, such as dispersion, on the Bogoliubov phases, and in \appref{4point}, we demonstrate how the four-point statistics of $\delrho(\p)$ can be used to distinguish a coherent state ensemble from a squeezed state if the coefficients are not in-phase.  This distinction relies on measuring a subdominant contribution to the four-point function, and so may be difficult in practice.

Without loss of generality, we may take $\alpha_\p,\beta_\p$ to be real for all $\p$.  If we make the additional assumption that the modes which are relevant in \eqref{2point_squeezed} have large particle number, then we may simply take $\alpha_\p\approx\beta_\p\approx\sqrt{E_\p P(\p)}$.  \eqref{2point_squeezed} then implies
\begin{equation}
    \ex{:\delrho(\p,t)\delrho(\q,t):}\Omega=2\ex{:\delrho(\p,t)\delrho(\q,t):}\Psi.
    \label{eq:2point_real}
\end{equation}

Note that this factor of two is only observable if we have fixed the amplitude of the power spectrum through another measurement, such as the observed mean energy density.  It is therefore useful to distinguish between two cases.  For a generic state, the mean energy density is given by
\begin{equation}
    \bar\rho\equiv\frac{\langle:\delrho(0,t):\rangle}{\delta^{(3)}(0)}=m^2\int d^3\k\,P_\mathrm{tot}(\k),
    \label{eq:rhobar}
\end{equation}
where $P_\mathrm{tot}(\k)$ is the total power spectrum of the scalar field, defined as in \eqref{trace_a1} for all modes $\k$, regardless of whether they are in a coherent state ensemble, squeezed state, or other state.  If the modes which dominate the integral in \eqref{rhobar} reside in a coherent state ensemble or squeezed state (as might be the case for vector DM produced during inflation~\cite{Graham:2015rva}), then $P_\mathrm{tot}(\k)$ may be identified with the power spectrum $P(\k)$ we have dealt with so far.  In this case, the amplitude of the power spectrum is fixed by observation of $\bar\rho$, and so the two-point statistics alone are enough to distinguish a coherent state ensemble from a squeezed state, via \eqref{2point_real}.

On the other hand, if the modes of $\hat\phi$ which reside in a coherent state ensemble or squeezed state are subdominant contributions to \eqref{rhobar}, then the amplitude of $P(\p)$ is not fixed.  This may be the case if DM is dominantly produced via another mechanism and inflation imprints subdominant isocurvature fluctuations on the DM.  In this case, another observable is required.  Below we show that, so long as the Bogoliubov coefficients are in-phase, the three-point function of $\delrho(\p)$ is sufficient.  In \appref{4point}, we show that if this is not the case, the four-point function may be able to distinguish a coherent state ensemble from a squeezed state.

\subsection{Three-point function}
\label{sec:3point}

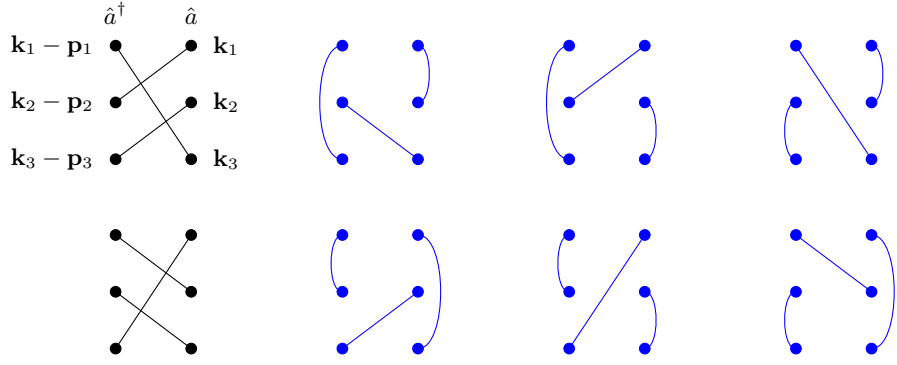
\begin{figure*}[t]
\begin{tikzpicture}
\tikzmath{\h=0.75;\w=1;\H=2.5;\W=3;\curve=0.2;}

\filldraw[black] (0,0) circle (2pt);
\filldraw[black] (0,\h) circle (2pt);
\filldraw[black] (0,2*\h) circle (2pt);
\filldraw[black] (\w,0) circle (2pt);
\filldraw[black] (\w,\h) circle (2pt);
\filldraw[black] (\w,2*\h) circle (2pt);
\draw[black] (0,0) -- (\w,2*\h);
\draw[black] (0,\h) -- (\w,0);
\draw[black] (0,2*\h) -- (\w,\h);

\filldraw[black] (0,\H) circle (2pt) node[left=5pt]{$\k_3-\p_3$};
\filldraw[black] (0,\h+\H) circle (2pt) node[left=5pt]{$\k_2-\p_2$};
\filldraw[black] (0,2*\h+\H) circle (2pt) node[above=5pt]{$\A^\dagger$} node[left=5pt]{$\k_1-\p_1$};
\filldraw[black] (\w,\H) circle (2pt) node[right=5pt]{$\k_3$};
\filldraw[black] (\w,\h+\H) circle (2pt) node[right=5pt]{$\k_2$};
\filldraw[black] (\w,2*\h+\H) circle (2pt) node[above=5pt]{$\A$} node[right=5pt]{$\k_1$};
\draw[black] (0,\H) -- (\w,\h+\H);
\draw[black] (0,\h+\H) -- (\w,2*\h+\H);
\draw[black] (0,2*\h+\H) -- (\w,\H);

\filldraw[blue] (\W,0) circle (2pt);
\filldraw[blue] (\W,\h) circle (2pt);
\filldraw[blue] (\W,2*\h) circle (2pt);
\filldraw[blue] (\w+\W,0) circle (2pt);
\filldraw[blue] (\w+\W,\h) circle (2pt);
\filldraw[blue] (\w+\W,2*\h) circle (2pt);
\draw[blue] (\W,0) -- (\w+\W,\h);
\draw[blue] (\W,\h) .. controls (\W-\curve,\h) and (\W-\curve,2*\h) .. (\W,2*\h);
\draw[blue] (\w+\W,0) .. controls (\w+\W+2*\curve,0) and (\w+\W+2*\curve,2*\h) .. (\w+\W,2*\h);

\filldraw[blue] (\W,\H) circle (2pt);
\filldraw[blue] (\W,\h+\H) circle (2pt);
\filldraw[blue] (\W,2*\h+\H) circle (2pt);
\filldraw[blue] (\w+\W,\H) circle (2pt);
\filldraw[blue] (\w+\W,\h+\H) circle (2pt);
\filldraw[blue] (\w+\W,2*\h+\H) circle (2pt);
\draw[blue] (\W,\h+\H) -- (\w+\W,\H);
\draw[blue] (\W,\H) .. controls (\W-2*\curve,\H) and (\W-2*\curve,2*\h+\H) .. (\W,2*\h+\H);
\draw[blue] (\w+\W,\h+\H) .. controls (\w+\W+\curve,\h+\H) and (\w+\W+\curve,2*\h+\H) .. (\w+\W,2*\h+\H);

\filldraw[blue] (2*\W,0) circle (2pt);
\filldraw[blue] (2*\W,\h) circle (2pt);
\filldraw[blue] (2*\W,2*\h) circle (2pt);
\filldraw[blue] (\w+2*\W,0) circle (2pt);
\filldraw[blue] (\w+2*\W,\h) circle (2pt);
\filldraw[blue] (\w+2*\W,2*\h) circle (2pt);
\draw[blue] (2*\W,0) -- (\w+2*\W,2*\h);
\draw[blue] (2*\W,\h) .. controls (2*\W-\curve,\h) and (2*\W-\curve,2*\h) .. (2*\W,2*\h);
\draw[blue] (\w+2*\W,0) .. controls (\w+2*\W+\curve,0) and (\w+2*\W+\curve,\h) .. (\w+2*\W,\h);

\filldraw[blue] (2*\W,\H) circle (2pt);
\filldraw[blue] (2*\W,\h+\H) circle (2pt);
\filldraw[blue] (2*\W,2*\h+\H) circle (2pt);
\filldraw[blue] (\w+2*\W,\H) circle (2pt);
\filldraw[blue] (\w+2*\W,\h+\H) circle (2pt);
\filldraw[blue] (\w+2*\W,2*\h+\H) circle (2pt);
\draw[blue] (2*\W,\h+\H) -- (\w+2*\W,2*\h+\H);
\draw[blue] (2*\W,\H) .. controls (2*\W-2*\curve,\H) and (2*\W-2*\curve,2*\h+\H) .. (2*\W,2*\h+\H);
\draw[blue] (\w+2*\W,\H) .. controls (\w+2*\W+\curve,\H) and (\w+2*\W+\curve,\h+\H) .. (\w+2*\W,\h+\H);

\filldraw[blue] (3*\W,0) circle (2pt);
\filldraw[blue] (3*\W,\h) circle (2pt);
\filldraw[blue] (3*\W,2*\h) circle (2pt);
\filldraw[blue] (\w+3*\W,0) circle (2pt);
\filldraw[blue] (\w+3*\W,\h) circle (2pt);
\filldraw[blue] (\w+3*\W,2*\h) circle (2pt);
\draw[blue] (3*\W,2*\h) -- (\w+3*\W,\h);
\draw[blue] (3*\W,0) .. controls (3*\W-\curve,0) and (3*\W-\curve,\h) .. (3*\W,\h);
\draw[blue] (\w+3*\W,0) .. controls (\w+3*\W+2*\curve,0) and (\w+3*\W+2*\curve,2*\h) .. (\w+3*\W,2*\h);

\filldraw[blue] (3*\W,\H) circle (2pt);
\filldraw[blue] (3*\W,\h+\H) circle (2pt);
\filldraw[blue] (3*\W,2*\h+\H) circle (2pt);
\filldraw[blue] (\w+3*\W,\H) circle (2pt);
\filldraw[blue] (\w+3*\W,\h+\H) circle (2pt);
\filldraw[blue] (\w+3*\W,2*\h+\H) circle (2pt);
\draw[blue] (3*\W,2*\h+\H) -- (\w+3*\W,\H);
\draw[blue] (3*\W,\H) .. controls (3*\W-\curve,\H) and (3*\W-\curve,\h+\H) .. (3*\W,\h+\H);
\draw[blue] (\w+3*\W,\h+\H) .. controls (\w+3*\W+\curve,\h+\H) and (\w+3*\W+\curve,2*\h+\H) .. (\w+3*\W,2*\h+\H);

\end{tikzpicture}
\caption{\label{fig:3point}%
    Contractions which contribute to the three-point function of the density perturbations [see \eqref[s]{3point_coherent} and (\ref{eq:3point_squeezed})].  Each diagram consists of six nodes representing the three creation and three annihilation operators in \eqref{3point_op} [with their corresponding momenta labeled in the top-left diagram], while lines indicate operator contractions.  As $\p_1,\p_2,\p_3\neq0$, horizontal contractions are forbidden.  There are two valid bipartite contractions, shown in black, which contribute to the result for the coherent state ensemble in \eqref{3point_coherent}.  Meanwhile, there are six other contractions, shown in blue, which contribute to the squeezed state result in \eqref{3point_squeezed}.  All diagrams in the top row yield the first term in \eqref{3point_coherent}, while the diagrams in the bottom row yield the second term.}
\end{figure*}

We now move on to calculate the three-point function of the density perturbations, in both an ensemble of coherent states and in a squeezed state.  The three-point density perturbation operator consists of three pairs of creation/annihilation operators.  As before, we consider all possible contractions of these six operators, which are shown in \figref{3point}.  In the case of the coherent state ensemble, we are again only interested in bipartite contractions.  As we are interested in $\p_1,\p_2,\p_3\neq0$, we cannot contract a creation operator with the annihilation operator from the same $\delrho(\p_i)$ operator.  This leaves only two possible contractions for the coherent state ensemble,
\begin{align}
    \label{eq:3point_op}
    \ex{:\delrho(\p_1)\delrho(\p_2)\delrho(\p_3):}\Psi&=m^3\int\frac{d^3\k_1}{(2\pi)^3}\frac{d^3\k_2}{(2\pi)^3}\frac{d^3\k_3}{(2\pi)^3}\ex{\A_{\k_1-\p_1}^\dagger\A_{\k_2-\p_2}^\dagger\A_{\k_3-\p_3}^\dagger\A_{\k_1}\A_{\k_2}\A_{\k_3}}\Psi\\
    &=\delta^{(3)}(\p_1+\p_2+\p_3)\cdot m^6\int d^3\k\,P(\k)P(\k-\p_1)\left[P(\k+\p_2)+P(\k+\p_3)\right].
    \label{eq:3point_coherent}
\end{align}

For the squeezed state case, we consider all contractions which do not pair $\A_{\k_i-\p_i}^\dagger$ with $\A_{\k_i}$.  \figref{3point} shows that this gives six additional contractions.  All of these contractions involve Bogoliubov coefficients which are not paired with their conjugates.  If the phase of the Bogoliubov coefficients vary with momentum, these contributions will be highly oscillatory and wash out.  In this case, we recover \eqref{3point_coherent}.  If, however, the Bogoliubov coefficients have constant phase, then these other six contractions contribute, and give a three-point function
\begin{equation}
    \ex{:\delrho(\p_1)\delrho(\p_2)\delrho(\p_3):}\Omega=4\ex{:\delrho(\p_1)\delrho(\p_2)\delrho(\p_3):}\Psi.
    \label{eq:3point_squeezed}
\end{equation}
Note the factor of four in \eqref{3point_squeezed}, in contrast to the factor of two in \eqref{2point_real}.  This cannot be simply accounted for by rescaling the power spectrum, and so this constitutes an observable difference between a coherent state ensemble and a squeezed state (in the case where the Bogoliubov coefficients have constant phase)!  This is a leading-order effect and persists even in the limit of large particle number.

Let us connect the quantities we have derived to some standard cosmological observables.  In cosmology, it is typical to deal with the density contrast $\hat\delta(\p)\equiv\frac{\delrho(\p)}{\bar\rho}$, where the mean energy density $\bar\rho$ is defined as in \eqref{rhobar}.  The matter power spectrum $P_\delta(\p)$ and matter bispectrum $B_\delta(\p_1,\p_2,\p_3)$ are then defined in terms of the statistics of $\hat\delta$
\begin{align}
    \langle:\hat\delta(\p)\hat\delta(\q):\rangle&\equiv(2\pi)^3P_\delta(\p)\delta^{(3)}(\p+\q),\\
    \langle:\hat\delta(\p_1)\hat\delta(\p_2)\hat\delta(\p_3):\rangle&\equiv(2\pi)^3B_\delta(\p_1,\p_2,\p_3)\delta^{(3)}(\p_1+\p_2+\p_3).
\end{align}
As discussed below \eqref{rhobar}, when the dominant fluctuations of DM are in a coherent state ensemble or squeezed state, $\bar\rho$ can be related to $P(\p)$.  In particular, suppose that the integral in \eqref{rhobar} is dominated by some scale $k_*$.  Then for $|\p|\gg k_*$, the matter power spectrum can be written as
\begin{equation}
    P_\delta(\p)\approx\frac{2m^2P(\p)}{(2\pi)^3\bar\rho},
\end{equation}
in the case of a coherent state ensemble.  Likewise, for $|\p_1|,|\p_2|,|\p_3|\gg k_*$, the bispectrum becomes
\begin{align}
    B_\delta(\p_1,\p_2,\p_3)&\approx\frac{2m^4[P(\p_1)P(\p_2)+P(\p_1)P(\p_3)+P(\p_2)P(\p_3)]}{(2\pi)^3\bar\rho^2}\\
    &=4\pi^3[P_\delta(\p_1)P_\delta(\p_2)+P_\delta(\p_1)P_\delta(\p_3)+P_\delta(\p_2)P_\delta(\p_3)]
    \label{eq:bispectrum_coherent}
\end{align}
for the coherent state ensemble.  In the typical language of non-Gaussianities~\cite{celoria2018}, this implies that the matter bispectrum exhibits a local shape for large $|\p_1|,|\p_2|,|\p_3|$, with $f_\mathrm{NL}=2\pi^3$, when the dominant DM fluctuations are described by an ensemble of coherent states.  It is not difficult to show that the squeezed state predicts the same $f_\mathrm{NL}$ (in the case of large $|\p_1|,|\p_2|,|\p_3|$).  Note that if subdominant fluctuations are described by one of these states, their $f_\mathrm{NL}$ can generically be much larger.

\section{Discussion}
\label{sec:discussion}

In this work, we computed the statistics of the density perturbations of a scalar field in two states.  The first is an ensemble of coherent states, which is often presumed for DM in laboratory or astrophysical contexts.  The second is a squeezed state, which is typically produced by inflation, or any other process which is described by a Bogoliubov transformation.  We showed that these two states can be distinguished solely based on observations of the matter power spectrum and bispectrum that they predict.  Importantly, these observations rely on no non-gravitational interactions of the DM.  This implies that cosmological observations of non-Gaussianity can act as probes of the current DM state, as well as its origin.

It is worth reiterating that both of the states considered in this work represent isocurvature perturbations of DM.  Adiabatic perturbations exhibit Gaussian statistics for $\hat\delta$ (in the linear regime of structure formation), and so the matter bispectrum predicted in \eqref{bispectrum_coherent} implies that these perturbations cannot be adiabatic.  The conclusion of this work can thus be phrased as follows: if large-scale isocurvature fluctuations of DM exist, they should exhibit observable non-Gaussianities, which may be able to elucidate their nature.  If DM is described by a scalar field, inflation would generically imprint isocurvature fluctuations onto it.  Moreover, observations of non-Gaussianities in the DM can probe smaller scales than direct observations of isocurvature.  Therefore, searching for non-Gaussianities may offer a promising method to diagnose isocurvature at certain scales.

Notably, our analysis has ignored dynamics of the DM, such as clustering.  We have assumed the current state of the DM to be a coherent state ensemble or squeezed state, and in the latter case, our results in \eqref[s]{2point_real} and (\ref{eq:3point_squeezed}) assumed phase coherence of the Bogoliubov coefficients.  Although inflation produces a squeezed state with phase coherence, it is far from clear that this state is not disrupted by clustering.  In \appref{dynamics}, we estimate the impact of the DM's dynamics on the state and show that they may be significant.  In \appref{4point}, we show that if the phase coherence is disrupted (but the squeezed state is preserved), a subdominant contribution to the four-point function of the density perturbations may be useful instead.  A more detailed treatment is required to understand how the statistics of the pertubations evolve through clustering.  Our work serves as a motivation for such future studies.

\acknowledgments

We thank Asimina Arvanitaki, Andrew Eberhardt, Peter Graham, Aurora Ireland, Nickolas Kokron, Zhen Liu, Nicholas Rodd, and Lian-Tao Wang for fruitful discussions on the topic of this work.  S.K. is supported in part by the DOE grant DE-SC0011842 and a Sloan Research Fellowship from the Alfred P. Sloan Foundation at the University of Minnesota.

\appendix

\section{Dynamics}
\label{app:dynamics}

In this appendix, we estimate the impact of dynamics on the squeezed state considered in this work.  Our calculations in \secref{perturbations} assumed not only that the current state of the DM is a squeezed state, but also that its Bogoliubov coefficients were in-phase.  As discussed below \eqref{2point_squeezed}, this will be the case initially for fluctuations which are produced by inflation, but clustering may affect either of these assumptions.  Here we estimate when either the phase coherence or the squeezed state itself is affected by dynamics.

We begin with the equation of motion for the scalar field in the presence of a gravitational potential $\Phi(\x,t)$.  The perturbed FRW metric with this potential (in Newtonian gauge) is given by
\begin{equation}
    ds^2=(1+2\Phi(\x,t))dt^2-a(t)^2(1-2\Phi(\x,t))|d\x|^2,
\end{equation}
where the scale factor $a(t)$ scales as
\begin{equation}
    a(t)\propto\left\{\begin{array}{cl}t^\frac12,&t<t_\mathrm{eq}\\t^\frac23,&t>t_\mathrm{eq}\end{array}\right.
\end{equation}
in radiation and matter domination, respectively, with current value $a(t_\mathrm{present})=1$.  In this metric, the Klein-Gordon equation for a scalar field $\hat\phi$ takes the form%
\footnote{In this appendix, for simplicity, we treat $\Phi$ as a classical background field.  During matter domination, $\Phi$ will be sourced by the DM.  Therefore, in reality, it should be treated as a dynamical quantum operator.  As the purpose of this appendix is to determine the regime in which the effects of $\Phi$ are negligible, we find it sufficient to ignore its dynamical and quantum nature.}
\begin{equation}
    \partial_t^2\hat\phi+(3H-4\partial_t\Phi)\partial_t\hat\phi-a^{-2}(1+4\Phi)\nabla^2\hat\phi+m^2(1+2\Phi)\hat\phi=0,
    \label{eq:klein-gordon}
\end{equation}
where $H=\frac{\partial_t a}a$.  When $\hat\phi$ is uniform in space, $\Phi$ is negligible, and $H\ll m$, this reduces to
\begin{equation}
    \partial_t^2\hat\phi+m^2\hat\phi=0,
    \label{eq:kg_simplified}
\end{equation}
so that each mode of $\hat\phi$ behaves as an independent harmonic oscillator with the same frequency $m$.  In this case, it is not difficult to see that if $\hat\phi$ begins in a squeezed state, it will remain in a squeezed state with Bogoliubov coefficients
\begin{align}
    \alpha_\p(t)&=\alpha_\p(0)e^{-imt},\\
    \beta_\p(t)&=\beta_\p(0)e^{-imt}.
\end{align}
This implies that if the coefficients begin in-phase, they will remain in-phase.  This squeezed state can be disrupted by either nonzero gradients of $\hat\phi$ or a nonzero potential $\Phi$.%
\footnote{The Hubble friction term $3H\partial_t\hat\phi$, on its own, does not disrupt the squeezed state.  This is because one can define $\hat\psi=a^{3/2}\hat\phi$, which satisfies \eqref{kg_simplified} to leading order.  Properly defining the state in terms of this operator makes it clear that the state remains squeezed.}
Specifically, the $a^{-2}\nabla^2\hat\phi$ term in \eqref{klein-gordon} will lead to finite dispersion, which can cause the various momentum modes to drift out of phase.  Likewise, the $m^2\Phi\hat\phi$ term can lead to mixing of different modes which can disrupt the squeezed state.

First let us analyze the impact of dispersion.  We can re-express $\hat\phi$ in terms of its Fourier components
\begin{equation}
    \hat\phi=\int\frac{d^3\p}{(2\pi)^3}\,\hat\phi_\p e^{i\p\cdot\x}.
\end{equation}
In terms of these Fourier modes, including the gradient term into \eqref{kg_simplified} yields
\begin{equation}
    \partial_t^2\hat\phi_\p+\left(\frac{|\p|^2}{a^2}+m^2\right)\hat\phi_\p=0.
\end{equation}
In this case, the Bogoliubov coefficients evolve with frequency $\tilde E_\p=\sqrt{\frac{|\p|^2}{a^2}+m^2}$.  The total phase accumulated by the coefficients of the mode with momentum $\p$ is%
\footnote{Here we assume $t_\mathrm{ent}(\p)<t_\mathrm{nr}(\p)<t_\mathrm{eq}$.  As DM must be non-relativistic by matter-radiation equality, the second equality is necessary.  For sufficiently large masses, \eqref{dispersion} allows for scales which have $t_\mathrm{ent}(\p)>t_\mathrm{nr}(\p)$.  In this case, the lower bound of the integral in \eqref{dispersion_phase} should really be $t_\mathrm{ent}(\p)$.  Even for masses as large as $m\sim1\,\mathrm{eV}$, this makes at most a $\sim10\%$ difference to the estimate in \eqref{dispersion}.}
\begin{equation}
    \int_{t_\mathrm{ent}(\p)}^{t_\mathrm{present}}dt\,\tilde E_\p\approx\int_{t_\mathrm{ent}(\p)}^{t_\mathrm{nr}(\p)}dt\,\frac{|\p|}a+\int_{t_\mathrm{nr}(\p)}^{t_\mathrm{eq}}dt\left(m+\frac{|\p|^2}{2a^2m}\right)+\int_{t_\mathrm{eq}}^{t_\mathrm{present}}dt\left(m+\frac{|\p|^2}{2a^2m}\right),
    \label{eq:dispersion_int}
\end{equation}
where $t_\mathrm{ent}(\p)$ is the time at which the mode enters the horizon, given by $|\p|=a(t_\mathrm{ent})H(t_\mathrm{ent})$; and $t_\mathrm{nr}(\p)$ is the time at which the mode becomes non-relativistic, given by $|\p|=a(t_\mathrm{nr})m$.  Note that the $m$ terms in the second two contributions to \eqref{dispersion_int} can be neglected, as they induce no dispersion.  Then, the first contribution is dominated by early times, while the last contribution is dominated by late times.  We therefore expect the dispersion to dominantly arise from the second term in \eqref{dispersion_int}.  This gives an accumulated relative phase
\begin{equation}
    \int_{t_\mathrm{nr}(\p)}^{t_\mathrm{eq}}dt\,\frac{|\p|^2}{2a^2m}\approx\frac{|\p|^2t_\mathrm{eq}}{a(t_\mathrm{eq})^2m}\log\frac{a(t_\mathrm{eq})m}{|\p|}.
    \label{eq:dispersion_phase}
\end{equation}
Requiring this phase to be less than one gives the constraint
\begin{equation}
    |\p|\ll30\,\mathrm{Mpc}^{-1}\left(\frac m{10^{-20}\,\mathrm{eV}}\right)^\frac12
    \label{eq:dispersion}
\end{equation}
(with an additional mild logarithmic dependence on $m$).

Now we estimate the impact of mode mixing induced by the gravitational potential on the squeezed state.  In terms of the Fourier modes of $\hat\phi$, including the dominant gravitational potential term into \eqref{kg_simplified} yields
\begin{equation}
    \partial_t^2\hat\phi_\p+m^2\hat\phi_\p+2m^2\int\frac{d^3\k}{(2\pi)^3}\,\Phi_{\p-\k}\hat\phi_\k=0,
    \label{eq:kg_potential}
\end{equation}
where $\Phi_\p$ are defined as
\begin{equation}
    \Phi=\int\frac{d^3\p}{(2\pi)^3}\,\Phi_\p e^{i\p\cdot\x}.
\end{equation}
As $\Phi\ll1$, \eqref{kg_potential} implies
\begin{equation}
    \partial_t\hat\phi_\p=\pm im\left(\hat\phi_\p+\int\frac{d^3\k}{(2\pi)^3}\,\Phi_{\p-\k}\hat\phi_\k\right),
    \label{eq:kg_firstorder}
\end{equation}
which gives two solutions for $\hat\phi_\p$ of (approximately) opposite frequencies.

Motivated by \eqref{scalar}, let us decompose $\hat\phi_\p$ into these two solutions as
\begin{equation}
    \hat\phi_\p=\frac{\A_\p(t)+\A_{-\p}^\dagger(t)}{\sqrt{2m}},
\end{equation}
with $\A_\p(t)\sim e^{-imt}$ the negative frequency solution and $\A_{-\p}^\dagger(t)\sim e^{imt}$ the positive frequency one.  Then \eqref{kg_firstorder} can be written as a differential equation for $\A_\p(t)$ as
\begin{equation}
    \partial_t\A_\p=-im\left(\A_\p+\int\frac{d^3\k}{(2\pi)^3}\,\Phi_{\p-\k}\A_\k\right).
\end{equation}
The general solution to this differential equation is
\begin{equation}
    \A_\p(t)=e^{-im(t-t_0)}\int\frac{d^3\k}{(2\pi)^3}\,U(\p,\k,t,t_0)\A_\k(t_0),
    \label{eq:amixing}
\end{equation}
where
\begin{equation}
    U(\p,\k,t,t_0)=(2\pi)^3\delta^{(3)}(\p-\k)-im\int_{t_0}^tdt_1\,\Phi_{\p-\k}(t_1)-m^2\int\frac{d^3\q}{(2\pi)^3}\int_{t_0}^tdt_1\int_{t_0}^{t_1}dt_2\,\Phi_{\p-\q}(t_1)\Phi_{\q-\k}(t_2)+\cdots.
\end{equation}
The solution for $\A_\p^\dagger$ is given by the complex conjugate of \eqref{amixing}.  Note that $U(\p,\k,t,t_0)^*=U(\k,\p,t_0,t)$.  We see that time evolution begins to mix the creation/annihilation operators of different modes.  The most noticeable effect of this mixing is that different modes will no longer be independent; more specifically, the correlators of creation/annihilation operators will no longer be proportional to delta functions, e.g.
\begin{align}
    \ex{\A_\p(t)\A_\q^\dagger(t)}\Omega&=\int\frac{d^3\k}{(2\pi)^3}\frac{d^3\k'}{(2\pi)^3}\,U(\p,\k,t,t_0)U(\q,\k',t,t_0)^*\ex{\A_\k(0)\A_{\k'}^\dagger(0)}\Omega\\
    &=\int\frac{d^3\k}{(2\pi)^3}\,|\alpha_\k|^2U(\p,\k,t,t_0)U(\k,\q,t_0,t)\\
    &=(2\pi)^3|\alpha_\p|^2\delta^{(3)}(\p-\q)-im\left(|\alpha_\p|^2-|\alpha_\q|^2\right)\int_{t_0}^tdt_1\Phi_{\p-\q}(t_1)+\cdots.
\end{align}
Note that the leading order mixing term is proportional to $|\alpha_\p|^2-|\alpha_\q|^2$, that is, modes with similar amplitude do not mix.  If we assume that the power spectrum $P(\p)$ is dominated by some scale $k_*$, then in order for a dominant mode $|\p|\sim k_*$ to mix, it must do so with a much larger or much smaller scale $|\q|\ll k_*$ or $|\q|\gg k_*$.  In other words, mixing requires nonzero $\Phi_\k$ for $|\k|\sim k_*$.  We can then estimate that mixing becomes relevant when
\begin{equation}
    m\cdot\frac{4\pi k_*^3}{3(2\pi)^3}\cdot\int_{t_\mathrm{ent}(k_*)}^{t_\mathrm{present}}dt\,\Phi_{k_*}(t)\sim1.
    \label{eq:mixing_int}
\end{equation}

In order to estimate the size of $\Phi_{k_*}$, we will apply standard linear cosmological perturbation theory~\cite{Baumann_2022}.  We note that the relevant modes which we consider in this appendix are small enough that nonlinear dynamics become important.  Therefore, this exercise only serves to give a rough idea of how large the mixing effects may be.  A full nonlinear computation is required to properly assess the impact of the gravitational potential.  In linear perturbation theory, $\Phi_\p$ begins outside the horizon with a scale-invariant spectrum%
\footnote{Note that if the DM isocurvature perturbations are large enough, the gravitational potential at the scale $k_*$ may be larger than the usual estimate from cosmological pertubation theory.  \eqref{cosmo_pt} should therefore be viewed as a lower bound on $\Phi_\p$.}
\begin{equation}
    \Phi_\p(t<t_\mathrm{ent}(\p))\sim4\times10^{-5}\cdot\frac{2\pi^2}{|\p|^3}.
    \label{eq:cosmo_pt}
\end{equation}
Once each mode enters the horizon, it begins to oscillate and decay in amplitude as $|\Phi_\p|\sim a^{-2}$ during radiation domination, and then remains constant during matter domination.  The resulting amplitude during matter domination is therefore
\begin{equation}
    \Phi_\p(t>t_\mathrm{eq})\sim4\times10^{-5}\cdot\frac{2\pi^2}{|\p|^3}\cdot\left\{\begin{array}{cl}1,&|\p|<k_\mathrm{eq}\\\left(\frac{a(t_\mathrm{eq})}{|\p|t_\mathrm{eq}}\right)^2,&|\p|>k_\mathrm{eq},\end{array}\right.
\end{equation}
where $k_\mathrm{eq}=a(t_\mathrm{eq})H(t_\mathrm{eq})\sim0.01\,\mathrm{Mpc}^{-1}$ is the scale that enters the horizon at matter-radiation equality.  We see that the integral in \eqref{mixing_int} is dominated by the period of matter (and dark-energy) domination.  In particular, it gives the constraint
\begin{equation}
    k_*\gg150\,\mathrm{Mpc}^{-1}\left(\frac m{10^{-20}\,\mathrm{eV}}\right)^\frac12.
    \label{eq:mixing}
\end{equation}
Comparing \eqref{dispersion} with \eqref{mixing}, we find that any dominant scale $k_*$ will be impacted by either dispersion or mode mixing from the gravitational potential.  However, the fact that these constraints are not too far apart suggests that parameter space may exist where both effects are nonzero, but do not completely destroy the squeezed state.  Curiously, we note that the inflationary production mechanism for vector DM proposed in \citeR{Graham:2015rva} predicts a dominant scale
\begin{equation}
    k_*\sim3\,\mathrm{mpc}^{-1}\left(\frac m{10^{-5}\,\mathrm{eV}}\right)^\frac12
\end{equation}
for DM masses $m\gtrsim10^{-5}\,\mathrm{eV}$, which lies between the bounds in \eqref[s]{dispersion} and (\ref{eq:mixing}).  [These scales, however, lie deep in the nonlinear regime of structure formation, so that \eqref{mixing} can not be directly applied.] 
 Finally, we comment that if \eqref{dispersion} is satisfied but \eqref{mixing} is not, the state will remain squeezed, but only lose its phase coherence.  In the next appendix, we show that the four-point statistics of the density perturbations may be useful in this case.  Either way, the estimates in this appendix motivate future work to precisely compute the dynamics of the squeezed state and its resulting perturbation statistics.

\section{Four-point function}
\label{app:4point}

\begin{figure*}[t]
\begin{tikzpicture}
\tikzmath{\h=0.65;\w=1;\H=2.75;\W=2.5;\curve=0.2;}

\filldraw[red] (-\W-\w/2,2*\H) circle (2pt);
\filldraw[red] (-\W-\w/2,\h+2*\H) circle (2pt);
\filldraw[red] (-\W-\w/2,2*\h+2*\H) circle (2pt);
\filldraw[red] (-\W-\w/2,3*\h+2*\H) circle (2pt);
\filldraw[red] (-\W+\w/2,2*\H) circle (2pt);
\filldraw[red] (-\W+\w/2,\h+2*\H) circle (2pt);
\filldraw[red] (-\W+\w/2,2*\h+2*\H) circle (2pt);
\filldraw[red] (-\W+\w/2,3*\h+2*\H) circle (2pt);
\draw[red] (-\W-\w/2,2*\H) -- (-\W+\w/2,\h+2*\H);
\draw[red] (-\W-\w/2,\h+2*\H) -- (-\W+\w/2,2*\H);
\draw[red] (-\W-\w/2,2*\h+2*\H) -- (-\W+\w/2,3*\h+2*\H);
\draw[red] (-\W-\w/2,3*\h+2*\H) -- (-\W+\w/2,2*\h+2*\H);

\filldraw[red] (-\w/2,2*\H) circle (2pt);
\filldraw[red] (-\w/2,\h+2*\H) circle (2pt);
\filldraw[red] (-\w/2,2*\h+2*\H) circle (2pt);
\filldraw[red] (-\w/2,3*\h+2*\H) circle (2pt);
\filldraw[red] (\w/2,2*\H) circle (2pt);
\filldraw[red] (\w/2,\h+2*\H) circle (2pt);
\filldraw[red] (\w/2,2*\h+2*\H) circle (2pt);
\filldraw[red] (\w/2,3*\h+2*\H) circle (2pt);
\draw[red] (-\w/2,2*\H) -- (\w/2,2*\h+2*\H);
\draw[red] (-\w/2,\h+2*\H) -- (\w/2,3*\h+2*\H);
\draw[red] (-\w/2,2*\h+2*\H) -- (\w/2,2*\H);
\draw[red] (-\w/2,3*\h+2*\H) -- (\w/2,\h+2*\H);

\filldraw[red] (\W-\w/2,2*\H) circle (2pt);
\filldraw[red] (\W-\w/2,\h+2*\H) circle (2pt);
\filldraw[red] (\W-\w/2,2*\h+2*\H) circle (2pt);
\filldraw[red] (\W-\w/2,3*\h+2*\H) circle (2pt);
\filldraw[red] (\W+\w/2,2*\H) circle (2pt);
\filldraw[red] (\W+\w/2,\h+2*\H) circle (2pt);
\filldraw[red] (\W+\w/2,2*\h+2*\H) circle (2pt);
\filldraw[red] (\W+\w/2,3*\h+2*\H) circle (2pt);
\draw[red] (\W-\w/2,2*\H) -- (\W+\w/2,3*\h+2*\H);
\draw[red] (\W-\w/2,\h+2*\H) -- (\W+\w/2,2*\h+2*\H);
\draw[red] (\W-\w/2,2*\h+2*\H) -- (\W+\w/2,\h+2*\H);
\draw[red] (\W-\w/2,3*\h+2*\H) -- (\W+\w/2,2*\H);

\filldraw[black] (-2*\W-\w/2,\H) circle (2pt) node[left=5pt]{$\k_4+\q$};
\filldraw[black] (-2*\W-\w/2,\h+\H) circle (2pt) node[left=5pt]{$\k_3-\q$};
\filldraw[black] (-2*\W-\w/2,2*\h+\H) circle (2pt) node[left=5pt]{$\k_2+\p$};
\filldraw[black] (-2*\W-\w/2,3*\h+\H) circle (2pt) node[above=5pt]{$\A^\dagger$} node[left=5pt]{$\k_1-\p$};
\filldraw[black] (-2*\W+\w/2,\H) circle (2pt) node[right=5pt]{$\k_4$};
\filldraw[black] (-2*\W+\w/2,\h+\H) circle (2pt) node[right=5pt]{$\k_3$};
\filldraw[black] (-2*\W+\w/2,2*\h+\H) circle (2pt) node[right=5pt]{$\k_2$};
\filldraw[black] (-2*\W+\w/2,3*\h+\H) circle (2pt) node[above=5pt]{$\A$} node[right=5pt]{$\k_1$};
\draw[black] (-2*\W-\w/2,\H) -- (-2*\W+\w/2,3*\h+\H);
\draw[black] (-2*\W-\w/2,\h+\H) -- (-2*\W+\w/2,\H);
\draw[black] (-2*\W-\w/2,2*\h+\H) -- (-2*\W+\w/2,\h+\H);
\draw[black] (-2*\W-\w/2,3*\h+\H) -- (-2*\W+\w/2,2*\h+\H);
\draw[black,dashed] (-2*\W-\w/2,\H) -- (-2*\W+\w/2,\h+\H);
\draw[black,dashed] (-2*\W-\w/2,2*\h+\H) -- (-2*\W+\w/2,3*\h+\H);

\filldraw[black] (-\W-\w/2,\H) circle (2pt);
\filldraw[black] (-\W-\w/2,\h+\H) circle (2pt);
\filldraw[black] (-\W-\w/2,2*\h+\H) circle (2pt);
\filldraw[black] (-\W-\w/2,3*\h+\H) circle (2pt);
\filldraw[black] (-\W+\w/2,\H) circle (2pt);
\filldraw[black] (-\W+\w/2,\h+\H) circle (2pt);
\filldraw[black] (-\W+\w/2,2*\h+\H) circle (2pt);
\filldraw[black] (-\W+\w/2,3*\h+\H) circle (2pt);
\draw[black] (-\W-\w/2,\H) -- (-\W+\w/2,\h+\H);
\draw[black] (-\W-\w/2,\h+\H) -- (-\W+\w/2,3*\h+\H);
\draw[black] (-\W-\w/2,2*\h+\H) -- (-\W+\w/2,\H);
\draw[black] (-\W-\w/2,3*\h+\H) -- (-\W+\w/2,2*\h+\H);
\draw[black,dashed] (-\W-\w/2,\h+\H) -- (-\W+\w/2,\H);
\draw[black,dashed] (-\W-\w/2,2*\h+\H) -- (-\W+\w/2,3*\h+\H);

\filldraw[black] (-\w/2,\H) circle (2pt);
\filldraw[black] (-\w/2,\h+\H) circle (2pt);
\filldraw[black] (-\w/2,2*\h+\H) circle (2pt);
\filldraw[black] (-\w/2,3*\h+\H) circle (2pt);
\filldraw[black] (\w/2,\H) circle (2pt);
\filldraw[black] (\w/2,\h+\H) circle (2pt);
\filldraw[black] (\w/2,2*\h+\H) circle (2pt);
\filldraw[black] (\w/2,3*\h+\H) circle (2pt);
\draw[black] (-\w/2,\H) -- (\w/2,2*\h+\H);
\draw[black] (-\w/2,\h+\H) -- (\w/2,\H);
\draw[black] (-\w/2,2*\h+\H) -- (\w/2,3*\h+\H);
\draw[black] (-\w/2,3*\h+\H) -- (\w/2,\h+\H);
\draw[black,dashed] (-\w/2,\H) -- (\w/2,\h+\H);
\draw[black,dashed] (-\w/2,3*\h+\H) -- (\w/2,2*\h+\H);

\filldraw[black] (\W-\w/2,\H) circle (2pt);
\filldraw[black] (\W-\w/2,\h+\H) circle (2pt);
\filldraw[black] (\W-\w/2,2*\h+\H) circle (2pt);
\filldraw[black] (\W-\w/2,3*\h+\H) circle (2pt);
\filldraw[black] (\W+\w/2,\H) circle (2pt);
\filldraw[black] (\W+\w/2,\h+\H) circle (2pt);
\filldraw[black] (\W+\w/2,2*\h+\H) circle (2pt);
\filldraw[black] (\W+\w/2,3*\h+\H) circle (2pt);
\draw[black] (\W-\w/2,\H) -- (\W+\w/2,\h+\H);
\draw[black] (\W-\w/2,\h+\H) -- (\W+\w/2,2*\h+\H);
\draw[black] (\W-\w/2,2*\h+\H) -- (\W+\w/2,3*\h+\H);
\draw[black] (\W-\w/2,3*\h+\H) -- (\W+\w/2,\H);
\draw[black,dashed] (\W-\w/2,\h+\H) -- (\W+\w/2,\H);
\draw[black,dashed] (\W-\w/2,3*\h+\H) -- (\W+\w/2,2*\h+\H);

\filldraw[black] (2*\W-\w/2,\H) circle (2pt);
\filldraw[black] (2*\W-\w/2,\h+\H) circle (2pt);
\filldraw[black] (2*\W-\w/2,2*\h+\H) circle (2pt);
\filldraw[black] (2*\W-\w/2,3*\h+\H) circle (2pt);
\filldraw[black] (2*\W+\w/2,\H) circle (2pt);
\filldraw[black] (2*\W+\w/2,\h+\H) circle (2pt);
\filldraw[black] (2*\W+\w/2,2*\h+\H) circle (2pt);
\filldraw[black] (2*\W+\w/2,3*\h+\H) circle (2pt);
\draw[black] (2*\W-\w/2,\H) -- (2*\W+\w/2,2*\h+\H);
\draw[black] (2*\W-\w/2,\h+\H) -- (2*\W+\w/2,3*\h+\H);
\draw[black] (2*\W-\w/2,2*\h+\H) -- (2*\W+\w/2,\h+\H);
\draw[black] (2*\W-\w/2,3*\h+\H) -- (2*\W+\w/2,\H);

\filldraw[blue] (-2*\W-\w/2,0) circle (2pt);
\filldraw[blue] (-2*\W-\w/2,\h) circle (2pt);
\filldraw[blue] (-2*\W-\w/2,2*\h) circle (2pt);
\filldraw[blue] (-2*\W-\w/2,3*\h) circle (2pt);
\filldraw[blue] (-2*\W+\w/2,0) circle (2pt);
\filldraw[blue] (-2*\W+\w/2,\h) circle (2pt);
\filldraw[blue] (-2*\W+\w/2,2*\h) circle (2pt);
\filldraw[blue] (-2*\W+\w/2,3*\h) circle (2pt);
\draw[blue] (-2*\W-\w/2,0) -- (-2*\W+\w/2,\h);
\draw[blue] (-2*\W-\w/2,3*\h) -- (-2*\W+\w/2,2*\h);
\draw[blue] (-2*\W-\w/2,\h) .. controls (-2*\W-\w/2-\curve,\h) and (-2*\W-\w/2-\curve,2*\h) .. (-2*\W-\w/2,2*\h);
\draw[blue] (-2*\W+\w/2,0) .. controls (-2*\W+\w/2+3*\curve,0) and (-2*\W+\w/2+3*\curve,3*\h) .. (-2*\W+\w/2,3*\h);
\draw[blue,dashed] (-2*\W-\w/2,\h) -- (-2*\W+\w/2,0);
\draw[blue,dashed] (-2*\W-\w/2,2*\h) -- (-2*\W+\w/2,3*\h);

\filldraw[blue] (-\W-\w/2,0) circle (2pt);
\filldraw[blue] (-\W-\w/2,\h) circle (2pt);
\filldraw[blue] (-\W-\w/2,2*\h) circle (2pt);
\filldraw[blue] (-\W-\w/2,3*\h) circle (2pt);
\filldraw[blue] (-\W+\w/2,0) circle (2pt);
\filldraw[blue] (-\W+\w/2,\h) circle (2pt);
\filldraw[blue] (-\W+\w/2,2*\h) circle (2pt);
\filldraw[blue] (-\W+\w/2,3*\h) circle (2pt);
\draw[blue] (-\W-\w/2,\h) -- (-\W+\w/2,0);
\draw[blue] (-\W-\w/2,3*\h) -- (-\W+\w/2,2*\h);
\draw[blue] (-\W-\w/2,0) .. controls (-\W-\w/2-2*\curve,0) and (-\W-\w/2-2*\curve,2*\h) .. (-\W-\w/2,2*\h);
\draw[blue] (-\W+\w/2,\h) .. controls (-\W+\w/2+2*\curve,\h) and (-\W+\w/2+2*\curve,3*\h) .. (-\W+\w/2,3*\h);
\draw[blue,dashed] (-\W-\w/2,0) -- (-\W+\w/2,\h);
\draw[blue,dashed] (-\W-\w/2,2*\h) -- (-\W+\w/2,3*\h);

\filldraw[blue] (-\w/2,0) circle (2pt);
\filldraw[blue] (-\w/2,\h) circle (2pt);
\filldraw[blue] (-\w/2,2*\h) circle (2pt);
\filldraw[blue] (-\w/2,3*\h) circle (2pt);
\filldraw[blue] (\w/2,0) circle (2pt);
\filldraw[blue] (\w/2,\h) circle (2pt);
\filldraw[blue] (\w/2,2*\h) circle (2pt);
\filldraw[blue] (\w/2,3*\h) circle (2pt);
\draw[blue] (-\w/2,0) -- (\w/2,\h);
\draw[blue] (-\w/2,2*\h) -- (\w/2,3*\h);
\draw[blue] (-\w/2,\h) .. controls (-\w/2-2*\curve,\h) and (-\w/2-2*\curve,3*\h) .. (-\w/2,3*\h);
\draw[blue] (\w/2,0) .. controls (\w/2+2*\curve,0) and (\w/2+2*\curve,2*\h) .. (\w/2,2*\h);
\draw[blue,dashed] (-\w/2,\h) -- (\w/2,0);
\draw[blue,dashed] (-\w/2,3*\h) -- (\w/2,2*\h);

\filldraw[blue] (\W-\w/2,0) circle (2pt);
\filldraw[blue] (\W-\w/2,\h) circle (2pt);
\filldraw[blue] (\W-\w/2,2*\h) circle (2pt);
\filldraw[blue] (\W-\w/2,3*\h) circle (2pt);
\filldraw[blue] (\W+\w/2,0) circle (2pt);
\filldraw[blue] (\W+\w/2,\h) circle (2pt);
\filldraw[blue] (\W+\w/2,2*\h) circle (2pt);
\filldraw[blue] (\W+\w/2,3*\h) circle (2pt);
\draw[blue] (\W-\w/2,\h) -- (\W+\w/2,0);
\draw[blue] (\W-\w/2,2*\h) -- (\W+\w/2,3*\h);
\draw[blue] (\W-\w/2,0) .. controls (\W-\w/2-3*\curve,0) and (\W-\w/2-3*\curve,3*\h) .. (\W-\w/2,3*\h);
\draw[blue] (\W+\w/2,\h) .. controls (\W+\w/2+\curve,\h) and (\W+\w/2+\curve,2*\h) .. (\W+\w/2,2*\h);
\draw[blue,dashed] (\W-\w/2,0) -- (\W+\w/2,\h);
\draw[blue,dashed] (\W-\w/2,3*\h) -- (\W+\w/2,2*\h);

\filldraw[black] (2*\W-\w/2,0) circle (2pt);
\filldraw[black] (2*\W-\w/2,\h) circle (2pt);
\filldraw[black] (2*\W-\w/2,2*\h) circle (2pt);
\filldraw[black] (2*\W-\w/2,3*\h) circle (2pt);
\filldraw[black] (2*\W+\w/2,0) circle (2pt);
\filldraw[black] (2*\W+\w/2,\h) circle (2pt);
\filldraw[black] (2*\W+\w/2,2*\h) circle (2pt);
\filldraw[black] (2*\W+\w/2,3*\h) circle (2pt);
\draw[black] (2*\W-\w/2,0) -- (2*\W+\w/2,3*\h);
\draw[black] (2*\W-\w/2,\h) -- (2*\W+\w/2,2*\h);
\draw[black] (2*\W-\w/2,2*\h) -- (2*\W+\w/2,0);
\draw[black] (2*\W-\w/2,3*\h) -- (2*\W+\w/2,\h);

\end{tikzpicture}
\caption{\label{fig:4point}%
    Contractions which contribute to the four-point function of the density perturbations.  As in \figref{3point}, each diagram consists of nodes corresponding to creation/annihilation operators, with contractions shown by solid lines.  In this figure, we set $\p=\p_1=-\p_2$ and $\q=\p_3=-\p_4$.  In red and black, we show all bipartite contractions, which contribute to the four-point function of the coherent state ensemble.  The red contractions are disconnected, and so do not contribute to $K_\Psi(\p,\q)$, as in \eqref{4point_def}.  [Only the leftmost is nonzero when $\p\neq\pm\q$.]  The blue diagrams show non-bipartite contractions which are phase-independent, and so contribute to $K_\Omega(\p,\q)$ for the squeezed state.  The dashed lines represent pairs of nodes which have not been contracted, but whose momenta are equal because we have fixed $\p_1=-\p_2$ or $\p_3=-\p_4$.  As a result, the non-bipartite contractions in the blue diagrams depend on the same momenta.  This leads to a cancellation of the Bogoliubov phases, as in \eqref{blue_diagram}.  The bottom two rows of diagrams are arranged so that diagrams in the same column yield the same term in \eqref{4point_squeezed}.}
\end{figure*}
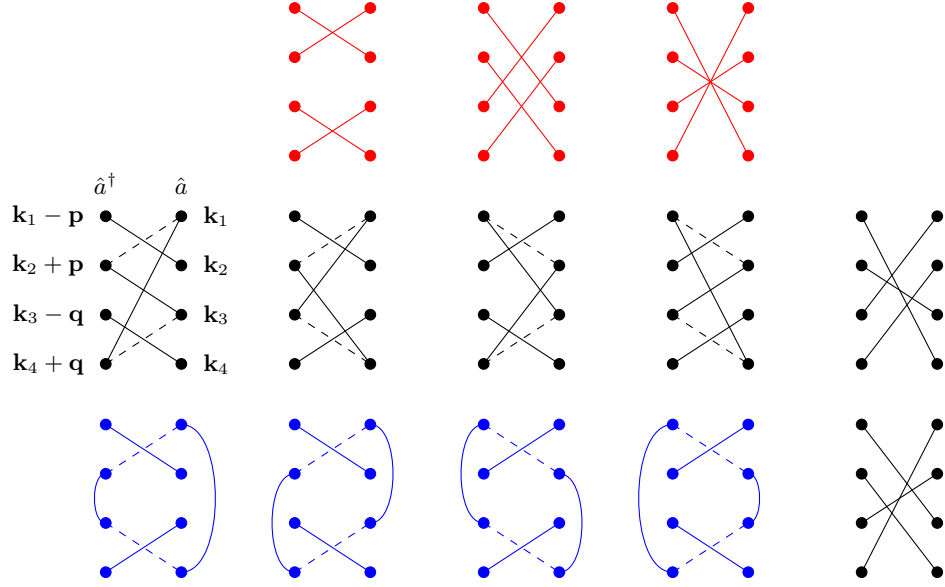

In the quantities we have computed so far, the case of a squeezed state with momentum-dependent phases for its Bogoliubov coefficients has been indistinguishable from a coherent state ensemble.  This is because contractions of two creation operators or two annihilation operators result in unpaired Bogoliubov coefficients, such as in \eqref{2point_squeezed}.  Generically, this will always be the case.  However, if we make an additional assumption about the relationship of the momenta entering the $n$-point correlator, the coefficients may incidentally become paired.  For instance, if we were to assume $\p=0$ in \eqref{2point_squeezed}, we would find that the integral only depends on the absolute values of the Bogoliubov coefficients and so will not be highly oscillatory.  As mentioned in \secref{perturbations}, we do not wish to set any momenta to zero, and so the two-point function cannot give us a nontrivial result while imposing this restriction.  For the three-point function, it also turns out that a nontrivial result would require setting one of the momenta to zero.

Therefore, we turn instead to the four-point function.  As we will see, if we set $\p_1=-\p_2$ and $\p_3=-\p_4$, then we will find a nonvanishing contribution to the squeezed state correlator, which does not appear for the coherent state ensemble.  As we have now assumed a relationship between the momenta, there will now be ``disconnected" contributions to the four-point function, i.e. contributions which come from the product of two two-point correlators.  We wish to isolate the connected contributions, that is, we will be interested in the quantity
\begin{equation}
    K_\Psi(\p,\q)\equiv\ex{\delrho(\p)\delrho(-\p)\delrho(\q)\delrho(-\q)}\Psi-\ex{\delrho(\p)\delrho(-\p)}\Psi\ex{\delrho(\q)\delrho(-\q)}\Psi,
    \label{eq:4point_def}
\end{equation}
for $\p,\q\neq0$ and $\p\neq\pm\q$, in the case of a coherent state ensemble.  We can similarly define $K_\Omega(\p,\q)$ for the squeezed state.

In total, there are nine bipartite contractions of the four-point density operator that do not pair creation and annihilation operators from the same $\delrho$ operator.  These are shown in red and black in \figref{4point}.  The three red diagrams constitute disconnected contributions.  (Only the leftmost is nonzero when $\p\neq\pm\q$.)  The other six contributions in black give
\begin{align}
    K_\Psi(\p,\q)=\delta^{(3)}(0)\cdot m^8\int d^3\k\,P(\k)P(\k-\p)&\Big[P(\k)P(\k-\q)+P(\k)P(\k+\q)+P(\k-\p)P(\k-\p-\q)\nl
    +P(\k-\p)P(\k-\p+\q)+2P(\k+\q)P(\k-\p+\q)\Big],
\label{eq:4point_coherent}
\end{align}
for $\p,\q\neq0$ and $\p\neq\pm\q$.

Now let us consider the squeezed state case.  Due to our choice of momenta, there are now non-bipartite contractions, which yield phase-independent contributions.  For instance, the leftmost blue diagram in \figref{4point} yields
\begin{align}
    K_\Omega(\p,\q)&\supset m^4\int\frac{d^3\k_1}{(2\pi)^3}\frac{d^3\k_2}{(2\pi)^3}\frac{d^3\k_3}{(2\pi)^3}\frac{d^3\k_4}{(2\pi)^3}\ex{\A_{\k_2+\p}^\dagger\A_{\k_3-\q}^\dagger}\Omega\ex{\A_{\k_1}\A_{\k_4}}\Omega\ex{\A_{\k_1-\p}^\dagger\A_{\k_2}}\Omega\ex{\A_{\k_4+\q}^\dagger\A_{\k_3}}\Omega\\
    &=\delta^{(3)}(0)\cdot m^4\int d^3\k_1\,\alpha^*_{-\k_1}\beta^*_{\k_1}\cdot\alpha_{\k_1}\beta_{-\k_1}\cdot|\beta_{\k_1-\p}|^2\cdot|\beta_{-\k_1+\q}|^2\label{eq:blue_diagram}\\
    &\approx\delta^{(3)}(0)\cdot m^8\int d^3\k\,P(\k)^2P(\k-\p)P(\k-\q),
\end{align}
where we have taken the limit of large particle number, so that $|\alpha_\k|^2\approx|\beta_\k|^2\approx E_\k P(\k)$.  Note that the phases of $\alpha_\k$ and $\beta_\k$ are irrelevant to this result, as only their absolute values appear.  In total, there are four of these non-bipartite phase-independent contractions, which are shown in blue in \figref{4point}.  In addition to the six black contractions, these yield a four-point function
\begin{align}
    K_\Omega(\p,\q)=2\delta^{(3)}(0)\cdot m^8\int d^3\k\,P(\k)P(\k-\p)&\Big[P(\k)P(\k-\q)+P(\k)P(\k+\q)+P(\k-\p)P(\k-\p-\q)\nl
    +P(\k-\p)P(\k-\p+\q)+P(\k+\q)P(\k-\p+\q)\Big].
    \label{eq:4point_squeezed}
\end{align}
for the squeezed state.  Note that \eqref{4point_squeezed} is not proportional to \eqref{4point_coherent}.  This implies that the connected four-point function for a squeezed state with momentum-dependent phases will have a different shape as well as normalization from a coherent state ensemble (as opposed to the three-point function in \eqref{3point_squeezed} which only differs from \eqref{3point_coherent} in normalization).  As before, the power spectrum $P(\k)$ can be inferred by observing the average energy density or two-point density correlator, and then the observed four-point function can be compared to \eqref[s]{4point_coherent} and (\ref{eq:4point_squeezed}) to determine the state of $\hat\phi$.  As a final comment, we note that the quantity $K(\p,\q)$ is subleading in the sense that the disconnected contributions which have been subtracted in \eqref{4point_def} scale as $\delta^{(6)}(0)$, so observing $K(\p,\q)$ requires precise cancellation of these contributions.

\bibliographystyle{JHEP}
\bibliography{references.bib}

\end{document}